# A Probabilistic Calibration Procedure for the CORSAIR Polarimeter

Alan Hsu[1,2] · Jenna Samra[1,2] · Steven Tomczyk[3] · Maxim Kramar[4,5]

**Abstract**
We present a novel Bayesian model and a corresponding robust, probabilistic calibration procedure for the CORSAIR polarimeter that can be applied to other polarimeters. Our calibration procedure combines existing Mueller matrix representations of polarimeters with Bayesian methods, and computes the posterior distribution of the parameters by collecting data from the polarimeter at different states. We show that the algorithm is able to converge and recover a well-constrained posterior of the free parameters with a credible interval that is consistent with the ground truth values. Posterior predictive checks indicate that our generative model with inferred parameters can reproduce the calibration data within the predictive uncertainty, and captures the dominant systematic effects of the calibration procedure. We further show that we can propagate calibration uncertainties in the distributions to downstream reconstructions of Stokes measurements and magnetic-field estimates. We find that the contribution of calibration uncertainty towards the reconstructed results is minimal relative to that of the photon noise uncertainty, indicating that estimates using our Bayesian calibration algorithm can achieve photon noise-limited measurements in the magnetic-field parameters. Finally, we test the Bayesian calibration algorithm on a lab prototype of the CORSAIR polarimeter, and show that it converges and closely recovers theoretical estimates of the free parameters from real-world measurements.

## 1. Introduction

Space-weather events, such as solar flares and coronal mass ejections, arise from magnetic-reconnection events in the solar atmosphere and can cause large disruptions to modern

✉ A. Hsu
alan.hsu@cfa.harvard.edu

J. Samra
jsamra@cfa.harvard.edu

1 Department of Astronomy, Harvard University, Cambridge, MA 02138, USA

2 Center for Astrophysics | Harvard & Smithsonian, Cambridge, MA 02138, USA

3 Solar Scientific LLC, Boulder, CO, USA

4 DKIST Science Support Center, National Solar Observatory, Pukalani, HI 96768, USA

5 Institute for Astronomy, University of Hawaii at Manoa, Pukalani, HI 96768, USA

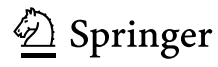

technological systems on Earth (e.g. Parker 1979; Golub and Pasachoff 2009; Aschwanden 2019). In order to understand and predict space weather, scientists model the magnetic fields that structure the corona and drive energy transport within it. Due to the challenge of measuring the coronal magnetic field, these models are generally constrained by photospheric magnetic fields and guided by coronal images (e.g. Mikić et al. 1999; Riley et al. 2006; van der Holst et al. 2014; Savcheva et al. 2015).

Spectropolarimeters can provide line-of-sight (LOS) intensity-weighted observations of the coronal magnetic field by measuring the Stokes polarization state of coronal emission lines (Judge et al. 2001). The linear polarization due to the Hanle Effect is captured in Stokes Q and U, while the circular polarization due to the Zeeman Effect is captured in Stokes V. Stokes Q and U can be used to constrain the vector field orientation and to recover an estimate of the magnetic azimuth under the plane-of-sky (POS) assumption, subject to the 90° ambiguity of the Van Vleck Effect. Stokes V can provide a measurement of the LOS component of the field strength (Judge et al. 2001; Casini and Judge 1999; Lin and Casini 2000). These Stokes vectors can also be used in single-point inversions to estimate magnetic orientation and thermal properties of the plasma (e.g. Paraschiv and Judge 2022; Judge, Casini, and Paraschiv 2021), coronal seismology to estimate global POS magnetic field values (e.g. Tomczyk and McIntosh 2009; Yang et al. 2020), and tomographic inversion techniques to estimate full vector coronal magnetic fields (e.g. Kramar, Inhester, and Solanki 2006; Kramar and Inhester 2007; Kramar et al. 2013; Kramar, Lin, and Tomczyk 2016; Kramar and Lin 2026).

The POS magnetic azimuth of the corona has been calculated regularly within the past two decades, such as from the Coronal Multichannel Polarimeter (CoMP, Tomczyk et al. 2008) and its upgrade UCoMP (Landi, Habbal, and Tomczyk 2016; Tomczyk and Landi 2019; Tomczyk et al. 2021) in the 1074.7 and 1079.8 nm Fe XIII lines. Deriving the coronal magnetic field strength from the faint Stokes V signal is much more challenging. A measurement of the LOS magnetic field strength from Stokes V (Schad et al. 2024) was recently obtained using the Cryogenic NIR Spectropolarimeter (Cryo-NIRSP, Fehlmann et al. 2023) of the Daniel K. Inouye Solar Telescope (DKIST, Rimmele et al. 2020). This measurement is the first since the technique was demonstrated (Lin, Penn, and Tomczyk 2000; Lin, Kuhn, and Coulter 2004) more than two decades ago. In the next few years, a new balloon-borne instrument called the Coronal Spectropolarimeter for Airborne Infrared Research (CORSAIR, Samra et al. 2021) will complement these ground-based observatories by providing uninterrupted, high-altitude, spectropolarimetric measurements of the coronal emission up to one solar radius from the disk. CORSAIR seeks to be the first to make continuous measurements of the global coronal Stokes V over a month-long solar rotation, a goal that poses significant technical challenges. Not only does the instrument have to achieve a particular sensitivity to detect Stokes V signals, but it also has to be calibrated accurately enough so that the uncertainty in the recontructed Stokes vectors is limited by photon noise and not calibration errors.

To measure the Stokes vectors, spectropolarimeters modulate the incoming intensity signal by changing the optical state of the polarimeter to encode the linear and circular polarization information in the intensity modulation, and recover the input Stokes vector by inverting the modulation process (e.g. del Toro Iniesta and Collados 2000). To achieve photon-noise-limited measurements of the Stokes vectors, polarimeter systematics must not limit the accuracy of the measured Stokes vectors. It is thus crucial to robustly calibrate the modulation sequence of the polarimeter.

Current spectropolarimeter calibration algorithms such as those for UCoMP (e.g. Tomczyk et al. 2008, Section 4) and Cryo-NIRSP (e.g. Harrington et al. 2023, Section 2.7) optimize an objective function to compute point-estimates of the best fitting parameters to the model. These methods cannot directly quantify the calibration uncertainty on these parameters, and can only compute them externally through other metrics. The UCoMP calibration procedure (see Appendix A) performs a Monte Carlo simulation on samples of measured Stokes vectors to compute a response matrix that characterizes how well the calibration performs on recovering the true Stokes vectors, and enforces that the calibration routine meet the error requirements on this matrix. Likewise, Cryo-NIRSP's calibration meets the error requirements based on an error matrix that is satisfied by fitting many parameters during the calibration sequence (e.g. Harrington et al. 2023, Section 2.8).

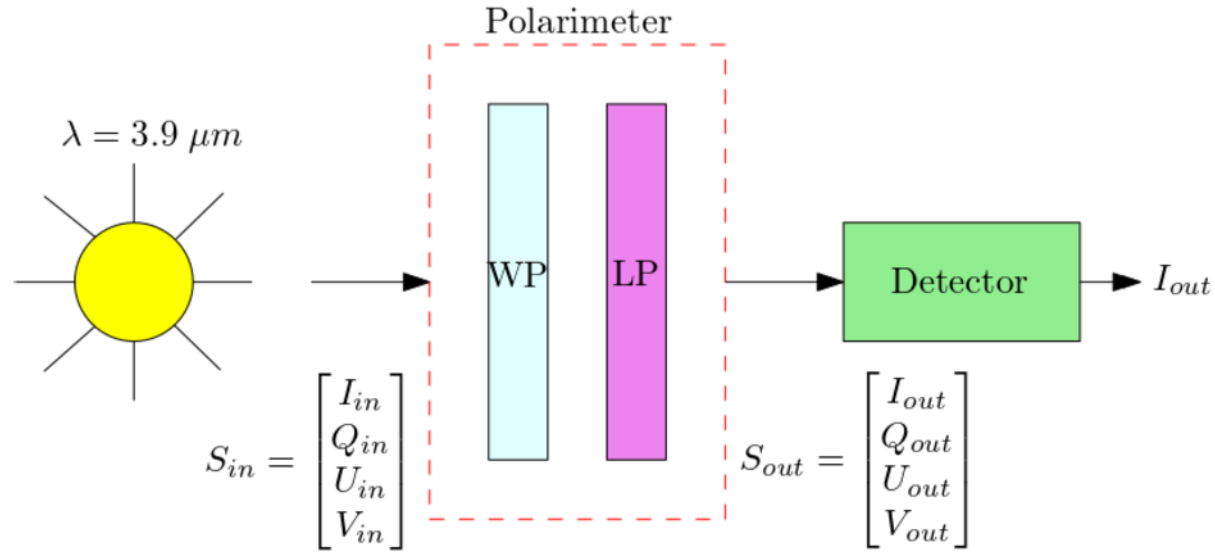

**Figure 1** Schematic Diagram of the CORSAIR Polarimeter. Infrared light from the solar corona has an incident polarization of $S_{\text{in}}$ (left) and is sent through the polarimeter with a known modulation scheme, consisting of a rotating waveplate (WP) and a fixed linear polarizer (LP). The detector measures the output modulated intensity $I_{\text{out}}$.

In this paper, we present a novel Bayesian model for spectropolarimeters and a corresponding probabilistic calibration procedure that computes the joint distribution of the parameters rather than point-estimates. Existing work in probabilistic calibrations of polarimeters has seen success in capturing the uncertainty of the calibration process through Mueller Matrix entries (e.g. Suárez-Bermejo et al. 2024). We extend on this work by predicting the uncertainty in the parameters of a Mueller Matrix model, and carry these distributions through downstream computations to explicitly quantify the contributions of calibration uncertainty towards the Stokes vectors and the magnetic field estimates. In Section 2, we provide a background on spectropolarimetric inversion and magnetic field estimates. In Section 3, we present our Bayesian model of the polarimeter and the calibration procedure to learn the joint distribution of the parameters. In Section 4, we show the robustness and accuracy of this calibration procedure using theoretical simulated data, and the propagation of the calibration uncertainties in downstream distributions of Stokes vectors and magnetic field estimates. Finally, in Section 5 we test the convergence and accuracy of the calibration procedure on a lab prototype of the CORSAIR polarimeter.

## 2. Coronal Spectropolarimetry

Measuring the full polarization state of light with photons is a challenging problem, since detectors can only directly measure intensity. Spectropolarimeters are designed to transform information of the polarization state of incident light into a modulated intensity output that detectors can measure: by inverting the modulation scheme, the incident polarization state can be recovered.

Figure 1 shows a schematic diagram of the CORSAIR Polarimeter, consisting of a waveplate (WP) and a linear polarizer (LP). The rotating waveplate modulates the polarization state of the incident light, injecting information of the linear and circular polarization into the output modulated intensity. The beam is then sent through a spectrograph (not shown) and then recorded on a detector array.

Assuming an ideal polarimeter and an ideal measurement setting, we can exactly model both the forward and inverse process using Mueller calculus (e.g. Goldstein and Chipman 1990; Tomczyk, Stoltz, and Seagraves 1991; del Toro Iniesta and Collados 2000; Tomczyk et al. 2010), where the polarization of light is a Stokes vector and optical elements are matrix operators. We can model the polarimeter matrix operator $P$ as a combination of the Mueller matrices of the WP and the LP:

$$S_{\mathrm{out}} = P(\theta, \phi) S_{\mathrm{in}} = M_{\mathrm{LP}} M_{\mathrm{WP}} S_{\mathrm{in}}, \tag{1}$$

where $\theta$ is a vector of angles that determine the orientation of the WP and LP, and $\phi$ is the retardance of the WP. Since the detector only measures $I_{\mathrm{out}}$, the first element of $S_{\mathrm{out}}$, only the first row of $P$ is relevant, denoted as $P_0$. As the waveplate $\theta$ rotates to discrete modulation states, the forward model $P_0(\theta, \phi)$ will change. Given $n$ modulation states $\{\theta_i\}_{i=1}^n$, we can construct an $n$ by 4 modulation matrix $\mathcal{M}$ that gives us the modulated intensity output by stacking $P_0$ at each state:

$$\begin{bmatrix} P_0(\theta_1) \\ P_0(\theta_2) \\ \vdots \\ P_0(\theta_n) \end{bmatrix} \times S_{\mathrm{in}} = \begin{bmatrix} I_{\mathrm{out},1} \\ I_{\mathrm{out},2} \\ \vdots \\ I_{\mathrm{out},n} \end{bmatrix} \tag{2}$$

$$\mathcal{M} \times S_{\mathrm{in}} = I_{\mathrm{meas}}. \tag{3}$$

This is exactly the forward model of the polarimeter over one modulation period that maps the input Stokes vector $S_{\mathrm{in}}$ to a vector of measured modulated intensities $I_{\mathrm{meas}}$. The inverse model is then the demodulation matrix $\mathcal{D}$ of the system which is the pseudoinverse of $\mathcal{M}$:

$$\tilde{S}_{\mathrm{in}} = \mathcal{D} I_{\mathrm{meas}} = (\mathcal{M}^T \mathcal{M})^{-1} \mathcal{M}^T I_{\mathrm{meas}}. \tag{4}$$

Since the spectrograph disperses the light onto the array, we can subsequently compute the incident polarization state as a function of wavelength, $S_{\mathrm{in}}(\lambda)$. The Stokes profiles can then be used to estimate the LOS strength and POS azimuth of the coronal magnetic field. The magnetic azimuth $\Phi_B$ can be computed using the wavelength-integrated linear polarization components $Q$ and $U$:

$$\Phi_B = \frac{1}{2} \arctan\left(\frac{U}{Q}\right). \tag{5}$$

The LOS field strength $B_{\mathrm{LOS}}$, in the weak field approximation, is a proportionality constant between the spectral intensity derivative and the wavelength-resolved circular polarization $V(\lambda)$:

$$V(\lambda) = \frac{\mu_B}{hc} \lambda^2 g_{\mathrm{eff}} \mathcal{K} B_{\mathrm{LOS}} \frac{\partial I}{\partial \lambda}, \tag{6}$$

where $\mu_B$ is the Bohr magneton, $h$ is Planck's constant, $c$ is the speed of light, $\lambda$ is the wavelength, $g_{\mathrm{eff}}$ is the effective Landè factor of the transition, and $\mathcal{K}$ is a correction for atomic alignment (Casini and Judge 1999; Lin and Casini 2000; Judge et al. 2001). We can solve for the best-fit $B_{\mathrm{LOS}}$ using linear least-squares.

## 3. Bayesian Calibration

### 3.1. Motivation

In practice, the exact form of the demodulation matrix $\mathcal{D}$ is unknown: instrument-specific systematics, optical imperfections such as partial polarization, birefringence, depolarization, and various sources of measurement and alignment noise can all cause the theoretical Mueller matrix model to deviate from the ground truth polarimeter. Thus, practical models of $\mathcal{D}$ have free parameters that are calibrated to capture all of these sources of deviations. Contrary to performing spectropolarimetry, where $\mathcal{D}$ is known and the modulated intensity signal is measured to recover the unknown incident Stokes vector $S_{\text{in}}$, calibrating a polarimeter requires a known Stokes vector input and measuring the output intensity to estimate $\mathcal{D}$. Generating a known Stokes vector input requires placing known polarization optics in front of the polarimeter, but in general these are not perfectly understood and need to be simultaneously calibrated.

Current standard techniques use numerical optimization subroutines to solve for a set of best parameters for $\mathcal{D}$ with respect to an objective function. For example, the UCoMP spectropolarimeter (Tomczyk et al. 2008) solves for the parameters of both the calibration optics in front of the polarimeter and the demodulation matrix by minimizing the L2-norm of the forward model Stokes vectors generated by the calibration optics and the inverse reconstructed Stokes vectors from a bootstrapped estimate of $\mathcal{D}$ (see Appendix A for more details). These techniques may give different solutions in the parameter space based on the stability of the numerical algorithms used or computational random variance. They also require external metrics post-calibration to quantify the uncertainty of the parameters and to validate the performance of $\mathcal{D}$. Finally, since $\mathcal{D}$ depends on the modulation scheme itself, the calibration output is limited to a specific modulation sequence that has to be enforced during spectropolarimetric measurements.

In this section, we will present a novel Bayesian model of the polarimeter and a corresponding statistical calibration algorithm that quantifies the calibration uncertainty of the solved parameters and propagates that to downstream estimates of the Stokes vectors and the magnetic fields. This method not only provides a robust convergence of parameters and their uncertainties, but is also independent of the modulation scheme used in downstream measurements.

### 3.2. Probabilistic Models for Polarimeters

Any existing parametric model of $\mathcal{D}$, defined by some set of free parameters $\boldsymbol{\omega}$, can be mapped to an analogous probabilistic model by replacing the parameters $\boldsymbol{\omega}$ with a joint probability distribution $p(\boldsymbol{\omega})$ over the same support: this new model uses an analogous distribution $p(\mathcal{D})$ instead of a point-estimate.

Existing calibration algorithms start the parameter search at some initial value $\boldsymbol{\omega}_0$ and use calibration data $C$ to solve for point-estimates $\boldsymbol{\omega}^*$ once the optimization routine converges, or equivalently, a shift $\Delta\boldsymbol{\omega} = \boldsymbol{\omega}^* - \boldsymbol{\omega}_0$. Analogously, calibrating a probabilistic model in the Bayesian formulation starts the parameters at some prior distribution $p_0(\boldsymbol{\omega})$, and solves for a distribution shift by computing the likelihood of the data collected $p(C|\boldsymbol{\omega})$. The posterior distribution $p(\boldsymbol{\omega}|C)$ on the parameters can then be computed from the likelihood as follows:

$$p(\boldsymbol{\omega}|C) \propto p(C|\boldsymbol{\omega})\, p_0(\boldsymbol{\omega}). \tag{7}$$

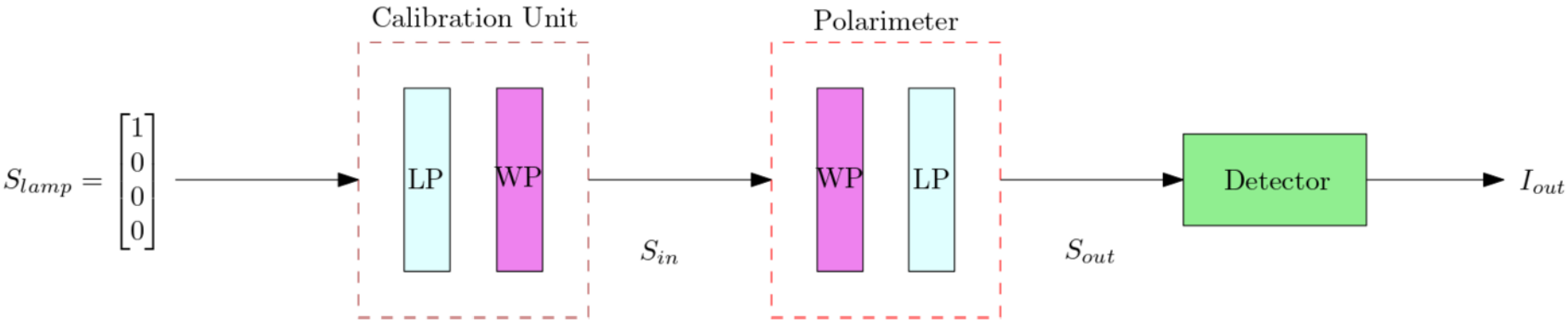

**Figure 2** Schematic Diagram of the calibration setup for the CORSAIR Polarimeter. Unpolarized light $S_{\text{lamp}}$ from a lamp is sent through calibration optics to produce a known Stokes vector $S_{\text{in}}$. The polarimeter is then modulated to produce an output intensity signal $I_{\text{out}}$, which is used to estimate the parameters $\omega$. In practice, $w$ may also encompass imperfections in $S_{\text{lamp}}$, such as deviations $dQ$ and $dU$ in Stokes $Q$ and $U$.

Figure 2 displays a schematic diagram of the CORSAIR polarimeter calibration. A calibration unit, consisting of a linear polarizer (LP) and a waveplate (WP), is placed in front of the polarimeter to generate known $S_{\text{in}}$ calibration Stokes vectors. The detector measures signal outputs $I_{\text{out}}$ corresponding to different polarimeter configurations, and solves for the most statistically likely distribution of the free parameters.

The schematic model can be mapped to a theoretical Mueller matrix forward model $M(\boldsymbol{\theta}, \boldsymbol{\omega})$, where the calibration unit (CU) and polarimeter (Pol) are represented by Mueller matrices that are functions of the orientation angle and free parameters to be calibrated, such as the birefringence, diattenuation, and physical offsets from the alignment procedures. The four orientation angles (two from CU and two from Pol) represent the state $\boldsymbol{\theta}$ of the system, while the free parameters are stored in the vector $\boldsymbol{\omega}$. In other words, our model $M$ maps a polarimeter configuration and a specific $\boldsymbol{\omega}$ to an output measured intensity. The Bayesian model is constructed under the assumption that the measured intensity is Gaussian about the forward model output:

$$I_i = M(\boldsymbol{\theta_i}, \boldsymbol{\omega}) + \sigma_I \xi_i, \tag{8}$$

where $I_i$ is the measured intensity from a specific model configuration $\boldsymbol{\theta_i}$, $\xi_i \sim \mathcal{N}(0, 1)$ is a normal random variable with mean 0 and variance 1, and $\sigma_I$ is a single, additional free parameter that represents the effective intensity measurement uncertainty.

The Gaussian noise model is justified by the high signal-to-noise ratio achievable during calibration, for which the true Poisson photon noise is well-approximated by a normal distribution. The $\sigma_I$ parameter is an effective noise parameter that absorbs photon noise and all unmodeled noise sources that are propagated through the nonlinear forward model $M(\boldsymbol{\theta}, \boldsymbol{\omega})$. Given the uniform exposure conditions and similar intensity scales across modulation states, a single $\sigma_I$ is the most natural choice that provides a concise and sufficient description of the residual intensity scatter. Appendix C demonstrates that in the limit of pure photon noise, the inferred $\sigma_I$ from our calibration accurately recovers the known noise level.

During calibration, we adhere to this forward model assumption and collect data by rotating the system to different configurations and measuring the output intensity. The collected data is thus a set of $N$ measurements $C = \{(I_i, \boldsymbol{\theta}_i)\}_{i=1}^{N}$. Assuming i.i.d. data, the likelihood is given by

$$P(C|\boldsymbol{\omega}, \sigma_I) = \prod_{i=1}^{N} \frac{1}{\sqrt{2\pi\sigma_I^2}} \exp\left[-\frac{(I_i - M(\boldsymbol{\theta}_i, \boldsymbol{\omega}))^2}{2\sigma_I^2}\right]. \tag{9}$$

We define all our free parameters in $\boldsymbol{\omega}$ as physical offsets and imperfections in the system, and are thus independent from each other (for example, the retardance of the polarimeter waveplate should not depend on that of the calibration waveplate). With these assumptions, the full log-posterior distribution of the free parameters post-calibration is given by

$$\log p(\boldsymbol{\omega}, \sigma_I | C) \sim \log p_0(\boldsymbol{\omega}, \sigma_I) - \frac{1}{2\sigma_I^2} \sum_{i=1}^{N} [I_i - M(\boldsymbol{\theta}_i, \boldsymbol{\omega})]^2 \,, \tag{10}$$

where $\log p_0(\boldsymbol{\omega}, \sigma_I)$ is a parameter-dependent log-prior that we set before the calibration.

### 3.3. Generative Model Summary

For clarity, we will summarize the generative model used throughout the work explicitly. For a polarimeter and calibration unit configuration $\boldsymbol{\theta}_i$ and set of free parameters $\boldsymbol{\omega}$, the measured intensity $I_i$ is generated by a deterministic Mueller-matrix forward model $M(\boldsymbol{\theta}_i, \boldsymbol{\omega})$ plus an additive noise term (Equation 8). We assume i.i.d. Gaussian noise with variance $\sigma_I^2$ for measurements, leading to the likelihood (Equation 9) and the posterior (Equation 10). Independent priors are placed on each physical parameter $\boldsymbol{\omega}$ and intensity uncertainty $\sigma_I$ (Table 2), with supports motivated by physical tolerances (Table 1).

### 3.4. Spectropolarimetric Bayesian Inference

Solving for the log-posterior of the free parameters during calibration provides a powerful tool for downstream inference of the Stokes vectors and the magnetic field. During observation, the calibration unit is removed and the data collected is a function of only the state of the polarimeter and the output modulated intensity. Given the modulation scheme of the observation, the log-posterior $\log p(\boldsymbol{\omega}, \sigma_I | C_{\text{cal}})$ can be mapped to a distribution of the demodulation matrix $\log p(\mathcal{D} | C_{cal})$. Equations 4–6 can then be used to produce posterior predictive distributions of the estimated Stokes vector and the magnetic field parameters $p(S_{\text{in}} | C_{\text{cal}}, C_{\text{obs}})$ and $p(B | C_{\text{cal}}, C_{\text{obs}})$, respectively. The distributions are conditioned on both the calibration data $C_{\text{cal}}$ as well as the observation data $C_{\text{obs}}$, with the parameter $\boldsymbol{\omega}$ dependency integrated out.

In practice, storing and using the full log-posterior of the parameters is unfeasible, largely due to the complex structure of the theoretical form and the high dimensional parameter space. Thus, we estimate the posterior distribution $p(\boldsymbol{\omega}, \sigma_I | C_{\text{cal}})$ defined by Equations 8-10 using Hamiltonian Monte Carlo (HMC) with the No-U-Turn Sampler (NUTS) to estimate the posterior. This is a gradient-based Markov chain Monte Carlo (MCMC) method that directly targets the posterior implied by our assumed Gaussian measurement model, sampling the posterior by simulating Hamiltonian dynamics on random walkers to produce representative samples of the landscape over large chains (e.g. Brooks et al. 2011; Hoffman and Gelman 2014). The NUTS algorithm alters the original HMC algorithm by introducing adaptive path lengths that terminate if the walker makes a U-turn: this not only removes the need to manually set a path length, but it also more efficiently explores the posterior geometry and improves the mixing and convergence of the sampler.

## 4. Simulation Results

### 4.1. CORSAIR Simulation Model

We validate our Bayesian calibration routine by simulating a model of CORSAIR and the data collection process. This allows us to specify and vary sources of noise, and compare the calibration results with ground truth parameters that would otherwise not be known in a real lab prototype.

**Table 1** Table of the free parameters $\boldsymbol{\omega}$ in our calibration model. We select physically-independent parameters and characterize imperfections on nominal values of these parameters. $dQ$ and $dU$ represent slight linear polarizations in the source lamp. $d\phi_{\mathrm{CU}}$ and $d\phi_{\mathrm{Pol}}$ represent deviations from expected retardance values of the Calibration Unit and Polarimeter waveplates. $d\theta_{\mathrm{CU}}$ and $d\theta_{\mathrm{Pol}}$ represent physical offsets from imperfect alignment of the waveplates. Ranges of tested values are listed in the rightmost column.

| Param | Description | Range |
|---|---|---|
| $dQ$ | Lamp $Q$ Imperfection | [0, 0.05] of Lamp $I$ |
| $dU$ | Lamp $U$ Imperfection | [0, 0.05] of Lamp $I$ |
| $d\phi_{\mathrm{CU}}$ | CU WP $\phi$ Imperfection | [0, 1] radians |
| $d\phi_{\mathrm{Pol}}$ | Pol WP $\phi$ Imperfection | [0, 1] radians |
| $d\theta_{\mathrm{CU}}$ | CU WP Offset | [0, 0.05] radians |
| $d\theta_{\mathrm{Pol}}$ | Pol WP Offset | [0, 0.05] radians |

The CORSAIR calibration setup is modeled using a Mueller calculus representation of Figure 2, with six free parameters ($\boldsymbol{\omega}$ is a vector of length 6) that capture various optical imperfections within the system, as described in Table 1. The parameters are selected such that they are physically independent of one another, and describe realistic imperfections within the system that are expected in a practical setup. We introduce slight linear polarization elements $dQ$ and $dU$ in the source lamp, imperfections $d\phi_{\mathrm{CU}}$, $d\phi_{\mathrm{Pol}}$ in the expected waveplate retardance angles of both the Calibration Unit and Polarimeter, and physical offsets $d\theta_{\mathrm{CU}}$, $d\theta_{\mathrm{Pol}}$ in the waveplates that arise from imperfect alignments. Because these parameters $\boldsymbol{\omega}$ are physically independent, they are separately identifiable during the calibration and result in meaningful estimates. In practice, adding more parameters will need careful verification: parameters that are highly correlated may result in data fits that are unphysical. For example, having multiple angle offsets as free parameters for both the LP and WP may be redundant and introduce symmetries that the data cannot break. Likewise, adding a lamp $V$ imperfection is unnecessary as the first CU LP is insensitive to the ciruclar polarization.

To simulate data collection, we set up a ground truth model of the Calibration Unit and Polarimeter with fixed parameters, and collect data by rotating the model to different states and recording the measured output intensity. While real-time usage of the polarimeter requires a specific modulation scheme, we choose to sample random states during calibration since we are estimating fundamental parameters and wish to break symmetries that may otherwise be hidden in particular modulation schemes. Realistic sources of Gaussian measurement noise in both the rotation angles of the optics and the measured intensity are added. The data is then fed into the calibration routine to compute a posterior distribution estimate of the parameters (the calibration routine has no access to the ground truth parameters.) The posterior distribution is saved via NUTS samples as implemented in the PyMC package (e.g. Abril-Pla et al. 2023), which are used to compute downstream estimates of Stokes and magnetic field parameters during simulated observations.

### 4.2. Posterior Distribution of Parameters

We test the Bayesian routine with our simulation pipeline using 50 calibration data samples. This number was selected based on numerical estimates of the MCMC posterior convergence (see Appendix B for more details). Figure 3 shows a corner plot of the posterior distribution of the parameters computed by the Bayesian calibration routine. We plot a kernel density estimate, or KDE, (blue) of the true distribution using a Gaussian smoothing kernel on a histogram of the MCMC samples. The diagonal shows the marginalized probability distribution for each parameter, while the off-diagonal plots are the pairwise 2-D projections of the parameters. We see that the ground truth values (dashed red) agree with the 94% highest density interval (HDI) spread (shaded blue) of the marginalized probabilities: the HDI is a Bayesian credible interval, stating that there is a 94% chance that the true parameter value lies within the interval after observing calibration data. It is the highest density interval in that all points outside the shaded region have a lower posterior probability density than those within. Because the ground truth values for all 6 parameters lie within the 94% HDI, it is a strong indication that the model learns a well-constrained posterior over the calibration parameters, with uncertainties consistent with ground truth. We also include a traditional point-estimate (dashed blue) for parameters predicted by the calibration model used by UCoMP, and we see close agreement between our recovered posterior and the baseline results. Because the baseline model only models and predicts the calibration optics and treats the polarimeter as a black box, only calibration optics parameters have a baseline estimate. Please refer to Appendix A for details on the baseline calibration algorithm.

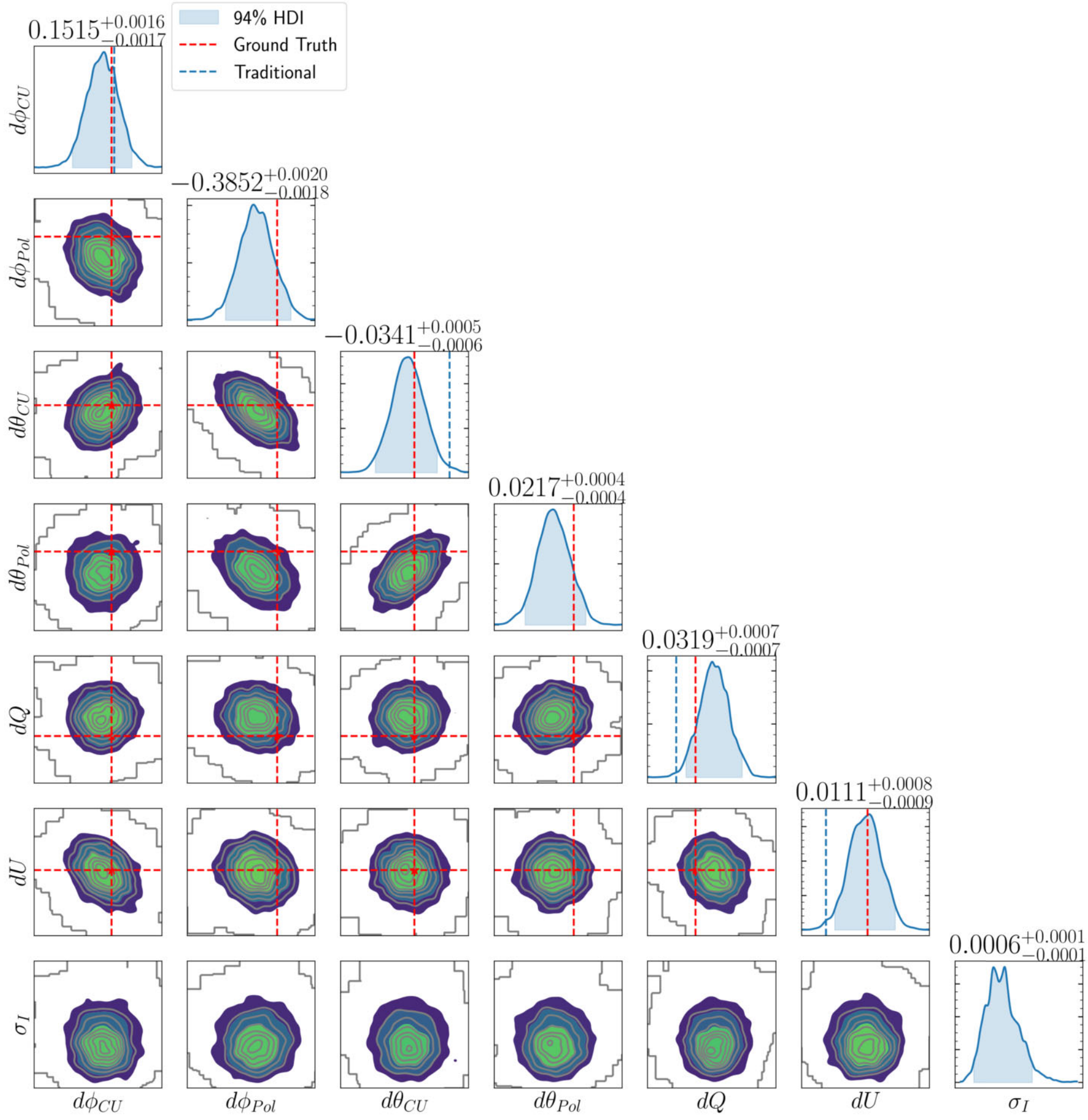

**Figure 3** Corner plot for a calibration experiment on the simulated CORSAIR model. The diagonal plots show a kernel density estimate of the marginalized posterior density of the 6 optical parameters $\boldsymbol{\omega}$ and the estimated intensity uncertainty $\sigma_I$, with the ground truth value (dashed red), the 94% highest density interval HDI (shaded blue), and the traditional point-estimate (dashed blue). The off-diagonal plots show the 2D projected posterior distributions, using Hamiltonian Monte Carlo NUTS samples to estimate the posterior density. We see that the calibration produces a well-constrained posterior over the parameters, with uncertainties consistent with ground truth. The calibrated posterior mean is also in close agreement to those predicted by the baseline non-probabilistic model. In addition, the minimal ellipticity of the projected densities indicates a largely independent parameter space.

**Table 2** Summary table of parameter values. The GT column is the ground truth value for the simulation model during data collection. The calibration model parameters are initialized to the prior given by the $p_0$ column, either $U(\min, \max)$ for a uniform or $HN(\mu, \sigma)$ for half-normal. The post calibration center and spread of the posterior is given by the $p_{\text{est}}$ column.

| Param | GT | $p_0$ | $p_{\text{est}}$ |
|---|---|---|---|
| $dQ$ | 0.0314 | $U(-0.1, 0.1)$ | $0.0319^{+0.0007}_{-0.0007}$ |
| $dU$ | 0.0112 | $U(-0.1, 0.1)$ | $0.0111^{+0.0008}_{-0.0009}$ |
| $d\phi_{\text{CU}}$ | 0.152 | $U(-1, 1)$ | $0.1515^{+0.0016}_{-0.0017}$ |
| $d\phi_{\text{Pol}}$ | −0.384 | $U(-1, 1)$ | $-0.3852^{+0.0020}_{-0.0018}$ |
| $d\theta_{\text{CU}}$ | −0.034 | $U(-0.1, 0.1)$ | $-0.0341^{+0.0005}_{-0.0006}$ |
| $d\theta_{\text{Pol}}$ | 0.022 | $U(-0.1, 0.1)$ | $0.0217^{+0.0004}_{-0.0004}$ |
| $\sigma_I$ | – | $HN(0, 0.05)$ | $0.0006^{+0.0001}_{-0.0001}$ |

Additionally, because the ground truth $\sigma_I$ is a complex combination of both angle uncertainty through the forward model and photon noise, we do not have a definitive ground truth, and treat the recovered $\sigma_I$ posterior as an additional output of our model. We also tested the simulated data with only photon noise, for which we have a definitive ground truth, and showed that the model could recover the value of $\sigma_I$. Please see Appendix C for a detailed discussion on this.

Table 2 summarizes the ground truth parameter values, the prior $p_0$, the posterior mean $p_{\text{est}}$, and 94% HDI values. The prior support is selected based on practical upper and lower bounds on these imperfections. All parameters have a uniform prior ($U$), with the exception of the $\sigma_I$ parameter which has a half-normal prior ($HN$). The ground truth values in the first column agree with the posterior mean and HDI interval in the last column. Given uniform distributions for these priors, the calibrated Gaussian-like peaks in posterior are an indication that the priors are not dominating, and the posterior is data-driven. With the peaks centered near the truths, we have a posterior that is locally identifiable, in that the calibration data is able to pin down a unique solution and break any degeneracies. In this regime, NUTS sampling works very effectively to quickly capture the posterior, and our Bayesian calibration routine becomes very robust.

### 4.3. Posterior Predictive Checks

Before propagating the calibrated posterior to downstream Stokes vectors and magnetic-field estimates, we verify that the inferred generative model reproduces the calibration data using posterior predictive checks (PPCs). To do so, we take the modulation states $\{\boldsymbol{\theta}_i\}_{i=1}^{N}$ of the calibration data, and reproduce a distribution of $I_{\mathrm{pred}}$ using MCMC samples of the posterior $p(\boldsymbol{\omega}, \sigma_I | C_{\mathrm{cal}})$, and see if it matches the observed data $\{I_i\}_{i=1}^{N}$.

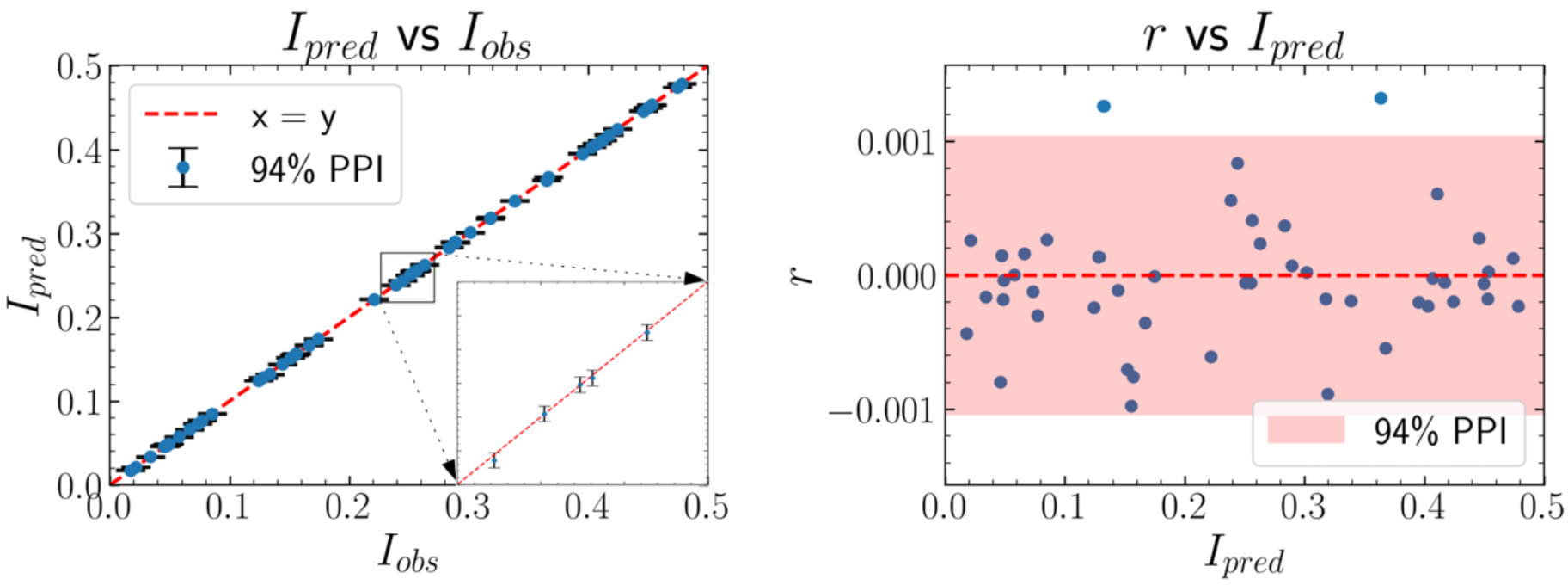

**Figure 4** Posterior predictive checks for the calibrated posterior. The left panel shows the reconstructed $I_{\mathrm{pred}}$ plotted against the calibration data $I_{\mathrm{obs}}$ for all 50 samples. The blue dot indicates the posterior predictive median, while the black bars indicate the 94% PPI. A magnified region is shown on the bottom left for visual clarity. The narrow fit indicates that the generative model with inferred parameters reproduces the calibration data within the expected predictive uncertainty. The right panel shows the residual plotted against $I_{\mathrm{pred}}$, with the 94% PPI shaded in red. The residuals show no discernible structure or intensity dependence, indicating that the calibration model learns the dominant systematic effects. The 2/50 samples outside the PPI are consistent with statistical expectation.

Figure 4 shows posterior predictive checks for the calibration intensities. The left panel shows the reconstructed $I_{\mathrm{pred}}$ plotted against the observed calibration data $I_{\mathrm{obs}}$ for all 50 calibration states, with the $I_{\mathrm{pred}} = I_{\mathrm{obs}}$ line dotted in red. Each point summarizes the full posterior predictive distribution (PPD) $p(I_{\mathrm{pred}}|C_{\mathrm{cal}})$ obtained by marginalizing over the calibrated parameter posterior. The blue marker indicates the posterior predictive median, and the black bars indicate the 94% posterior predictive interval (PPI), computed as the 94% HDI of the PPD. The PPIs are narrow, which is expected as it reflects the high SNR used in the simulated calibration measurements. A magnified portion of the plot (around $0.24 < I_{\mathrm{pred}} < 0.27$) is shown on the bottom right to make the PPIs more visible. Overall, the close agreement between $I_{\mathrm{pred}}$ and $I_{\mathrm{obs}}$ demonstrates that the generative model with the inferred parameters reproduces the calibration data within the expected predictive uncertainty.

The right panel shows the residuals $r = I_{\mathrm{obs}} - I_{\mathrm{pred}}$ plotted against $I_{\mathrm{pred}}$. Under the assumed generative model, the residuals should be consistent with unstructured Gaussian noise if the calibrated model accounts for all the systematic effects. Indeed, we observed no structure or intensity dependence in the residuals, indicating that the calibration model captures the dominant systematic effects. The region shaded in red is the 94% PPI of the residuals computed from the inferred noise parameter $\sigma_I$: out of 50 samples, two fall outside this interval, consistent with statistical expectations of three ($50 \times (1 - 0.94) = 3$). Together, these two diagnostics further confirm that the calibrated posterior provides a good description of the calibration data. See Appendix D for more posterior predictive checks on a larger calibration dataset.

We then compute two downstream metrics that further validate the reconstruction performance of the Bayesian calibration procedure. First, we compute the log-posterior of the demodulation matrix $\log P(\mathcal{D}|C_{\mathrm{cal}})$ for a uniformly rotating modulation scheme sampling measurements at $\pi/8$ intervals (the CORSAIR modulation scheme). In Figure 5, we show the posterior distributions for all the entries of the demodulation matrix, approximated by a Gaussian kernel density estimate (blue) and the corresponding 94% HDI (shaded blue).

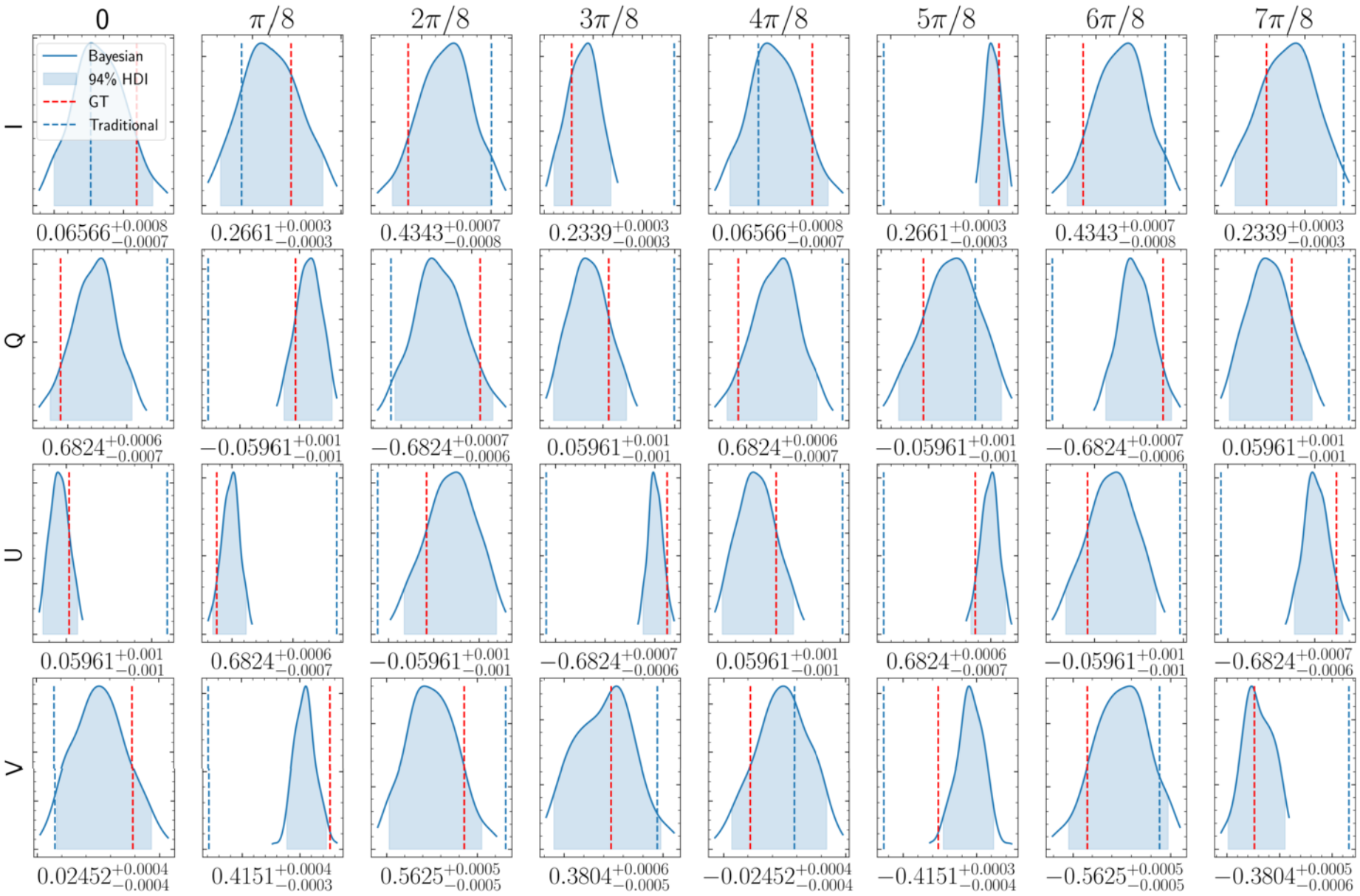

**Figure 5** Posterior predictive distribution of the demodulation matrix recovered from the Bayesian calibration. Each matrix entry shows a kernel density estimate of the posterior (blue) using the MCMC samples of the log-posterior, along with the 94% HDI interval (shaded blue), computed for a modulation scheme of eight evenly-spaced intervals from 0 to $\pi$. The ground truth (dashed red) and the traditional point-estimate (dashed blue) are plotted in vertical lines. We see that across the 32 entries, 30 of them (approximately 94% as expected) have the ground truth within the HDI, indicating that our Bayesian calibration algorithm captures the spread of the values due to calibration uncertainties and offers more information than a point estimate.

Alongside the KDE we also show the ground truth value (dashed red) as well as the baseline point estimate (dashed blue). On average, we find an RMS Z-score value of 1.298 across all 32 matrix entries, indicating that our model is slightly under-confident and thus conservative, and the posterior can be made tighter with more samples or a more optimized sampling procedure. However, we find that exactly 2 out of the 32 (93.75%) ground truth values fall out of the 94% credible interval, which is a very strong indication that our model is capturing the true value at the expected frequency.

Second, we compute the log-posterior of the error matrix $\log P(\mathcal{DM}^*|C_{\mathrm{cal}})$, which gives the distribution of an operator that maps an incident Stokes vector $S_{\mathrm{in}}$ onto the Polarimeter to the reconstructed Stokes vector $S_{\mathrm{est}}$: $\mathcal{DM}^* S_{\mathrm{in}} = S_{\mathrm{est}}$. This operator is computed by taking the experimentally recovered demodulation matrix $\mathcal{D}$ and multiplying it by the ground truth modulation matrix $\mathcal{M}^*$. In practice, to estimate this posterior distribution we take each of the MCMC samples of $\mathcal{D}$ and multiply it by $\mathcal{M}^*$. In Table 3, we show the posterior mean of the error matrix entries with the 94% HDI uncertainties. For an ideal calibration, the error matrix should be the identity matrix. The rows indicate the fractional cross-talk of the input Stokes vectors into the elements of the reconstructed Stokes vectors. Our recovered error matrix has values close to one on the diagonal and close to zero off the diagonal, indicating that we have a calibration that is very close to ideal.

**Table 3** Posterior mean values of error matrix entries and 94% HDI uncertainties from the Bayesian calibration. $\mathcal{O}(\epsilon)$ indicates machine precision. We see that the diagonal terms are close to one, and off-diagonal terms are close to zero, indicating good Stokes vector reconstruction capabilities of the calibrated posterior distribution. We can interpret a particular row as the fractional contribution of the input Stokes vector to the output Stokes element of that row. We see that the cross talks on the bottom row are negligible, which is particularly important as the $V$ signal is much smaller than the $Q$ and $U$ signals.

| | $I_{\text{in}}$ | $Q_{\text{in}}$ | $U_{\text{in}}$ | $V_{\text{in}}$ |
|---|---|---|---|---|
| $I_{\text{out}}$ | $1 \pm \mathcal{O}(\epsilon)$ | $-0.0007^{+0.0011}_{-0.0010}$ | $-0.0002^{+0.0004}_{-0.0004}$ | $\mathcal{O}(\epsilon)$ |
| $Q_{\text{out}}$ | $\mathcal{O}(\epsilon)$ | $1.0010^{+0.0010}_{-0.0011}$ | $0.0009^{+0.0014}_{-0.0015}$ | $\mathcal{O}(\epsilon)$ |
| $U_{\text{out}}$ | $\mathcal{O}(\epsilon)$ | $0.0009^{+0.0015}_{-0.0014}$ | $1.0010^{+0.0010}_{-0.0011}$ | $\mathcal{O}(\epsilon)$ |
| $V_{\text{out}}$ | $\mathcal{O}(\epsilon)$ | $\mathcal{O}(\epsilon)$ | $\mathcal{O}(\epsilon)$ | $0.9994^{+0.0010}_{-0.0008}$ |

### 4.4. Uncertainty Quantification of the Stokes Vectors and Magnetic Field Measurements

We assess the downstream performance of the Bayesian calibration results by testing the method on simulated spectropolarimetric measurements to reconstruct Stokes and magnetic field values. Our data consist of LOS-integrated Stokes vectors simulated using the second version of the Coronal Line Emission (CLE) code (e.g. Judge and Casini 2001; Judge, Low, and Casini 2006; Judge 2007). This code ingests 3D distributions of electron density, temperature, magnetic field, etc. and uses atomic data from CHIANTI (e.g. Dere et al. 1997, 2019) to generate LOS-integrated full Stokes vector observations for coronal emission lines. The 3D data is provided by the Magnetohydrodynamics Around a Sphere (MAS) model from Predictive Science, Inc. (PSI) (e.g. Mikić et al. 2007; Lionello, Linker, and Mikić 2008).

The first row of Figure 6 shows the spectropolarimetric data synthesized using CLE 2.0 and fed into our calibration model. The second row shows the Stokes vectors as a function of the wavelength at one particular pixel of the image. This dataset cube $S_{GT}$ spans $400 \times 400$ pixels over $4 \times 4$ solar radii in the spatial domain and 15 pixels in the spectral domain, enough to cover the line profile out to $\sim 3\sigma$.

To test our calibration, we send these measurements pixel-wise through the simulated polarimeter with ground truth calibration parameters to simulate the modulated intensity observations $\mathcal{C}_{obs}$. The observation exposure rates, which set the Poisson photon noise, are 30 seconds for $I$, 10 minutes for $Q$ and $U$, and 1 hour for $V$. These exposure times are selected based on experimental estimates of the SNR strength needed to reconstruct downstream magnetic field estimates, and what we expect to need for actual CORSAIR data. We then take our Bayesian calibration model and use the log-posterior distribution of the demodulation matrix $\log P(\mathcal{D}|\mathcal{C}_{cal})$ (see Figure 5) to invert the measured modulated intensity to recover the log-posterior of the input Stokes vector $\log P(S_{est}|\mathcal{C}_{cal}, \mathcal{C}_{obs})$. We then compute statistical metrics on this distribution and compare our results with $S_{GT}$. In practice, the computation of these log-posteriors is done using MCMC samples from the calibration.

Figure 7 shows sample reconstruction images of the PSI dataset integrated over wavelength. The plotted values are globally normalized such that the maximum intensity in the ground truth is unity for visual clarity. The first row shows the ground truth PSI dataset $S_{GT}$. The second row shows the statistical mean values $\mu_\gamma$ across 160 noisy samples of modulated intensity observation $\mathcal{C}_{obs}$: this can be interpreted as a perfect calibration where the posterior distribution is a Dirac delta distribution on the ground truth values, and the only noise source is due to the Poisson photon noise. The third row shows the statistical mean values $\mu_c$ across 160 MCMC samples of the calibration parameter posterior: this is the error due to systematics captured by the Bayesian calibration algorithm, independent of photon noise.

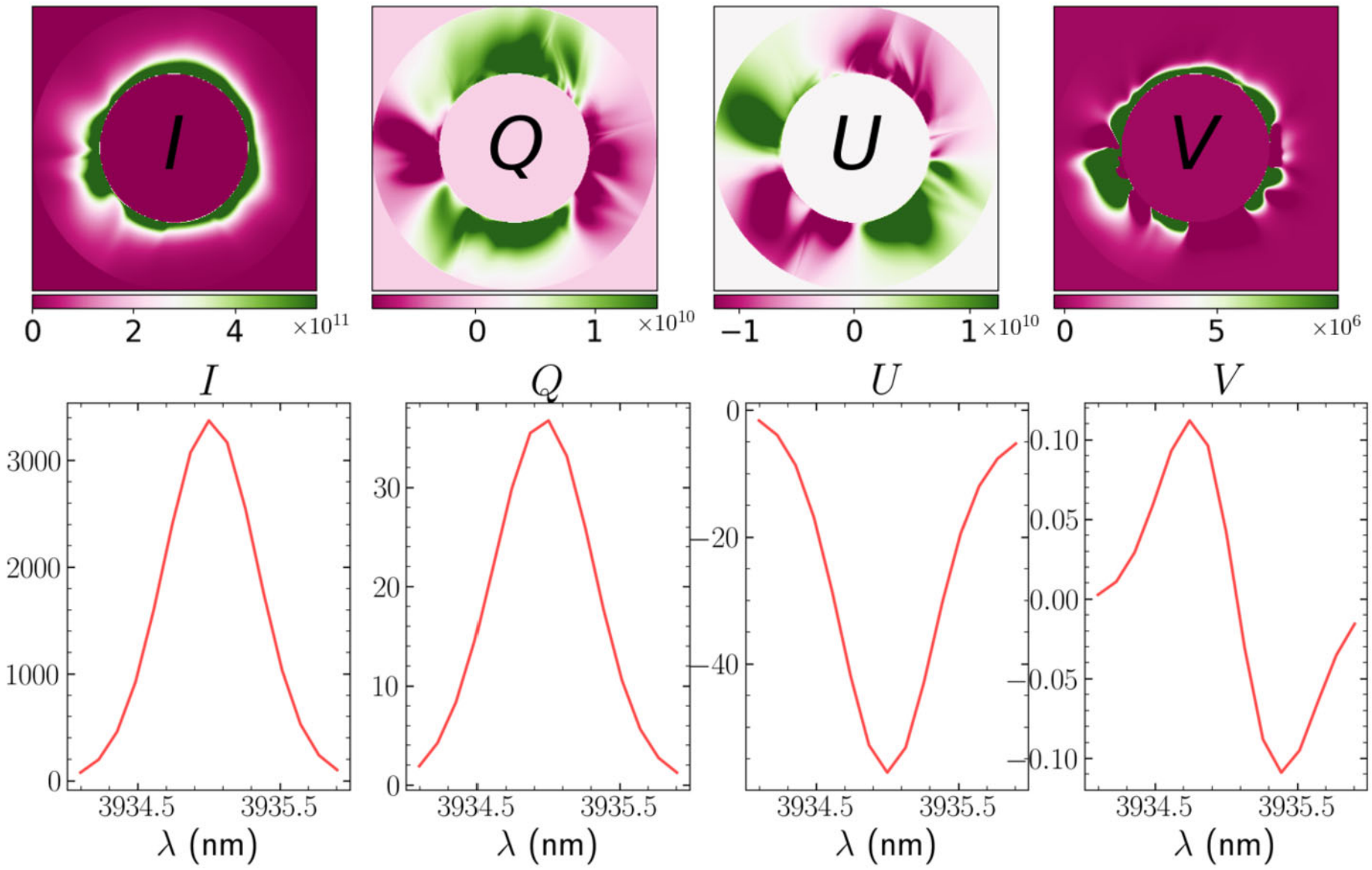

**Figure 6** Spectropolarimetric measurements generated from the PSI MAS model for Carrington Rotation 2149, which took place near solar maximum in April 2014. The first row shows LOS and wavelength integrated images of spectropolarimetric measurements, in photons s$^{-1}$ cm$^{-2}$ ster$^{-1}$, in the 3.9 µm Si IX emission line generated using CLE 2.0. The second row shows the Stokes vectors as a function of wavelength at a particular pixel in photons s$^{-1}$, assuming the effective area and solid angle of CORSAIR. We test the reconstruction of these synthesized spectropolarimetric observations using our calibrated polarimeter.

The fourth and fifth rows show the mean squared error due to photon noise ($MSE_\gamma$) and calibration error ($MSE_c$), respectively. The MSE is a good performance proxy for the reconstruction of Stokes vectors and magnetic field estimates because it is an easy-to-interpret scalar value that can incorporate different noise sources used in the simulations. Because our reconstruction method is not an unbiased estimator of the Stokes vector nor the magnetic field, for a given noise source we need to account for both the variance and bias. Using calibration error as an example, the overall mean-squared error for the calibration noise reconstruction is given by $MSE_c = \sigma_c^2 + b_c^2$, where $\sigma_c^2$ is the contribution of the variance from the samples, computed over the MCMC samples, and $b_c^2$ is the mean-squared bias given by $b_c^2 = (GT - \mu_c)^2$. By displaying the MSE on the same color scale, we see that across all four Stokes elements, the photon noise MSE is on average larger than the calibration error MSE, indicating our calibration noise will not dominate the photon measurements.

We can further test the reconstruction of magnetic field LOS intensity and azimuth by comparing our estimated posterior distributions of these parameters with ground truth values from the input dataset. Using Equations 5 and 6, we can map Stokes vector measurements to the LOS field strength $B_{\mathrm{LOS}}$ and the POS magnetic azimuth $\Phi_B$. Figure 8 shows reconstructed images of $B_{\mathrm{LOS}}$ and $\Phi_B$. The first row $GT$ shows the ground truth values computed directly from the ground truth stokes vector dataset $S_{GT}$ (first row of Figure 7). The second ($\mu_\gamma$) and third ($\mu_c$) rows show the mean reconstruction due to photon noise and calibration error, respectively. The fourth and fifth rows show the photon noise and calibration error MSE, respectively. Similar to Figure 7, we find that for both $B_{\text{LOS}}$ and $\Phi_B$, the calibration error MSE is smaller, on average, than the photon noise error, indicating that our measurements will not be calibration noise dominated.

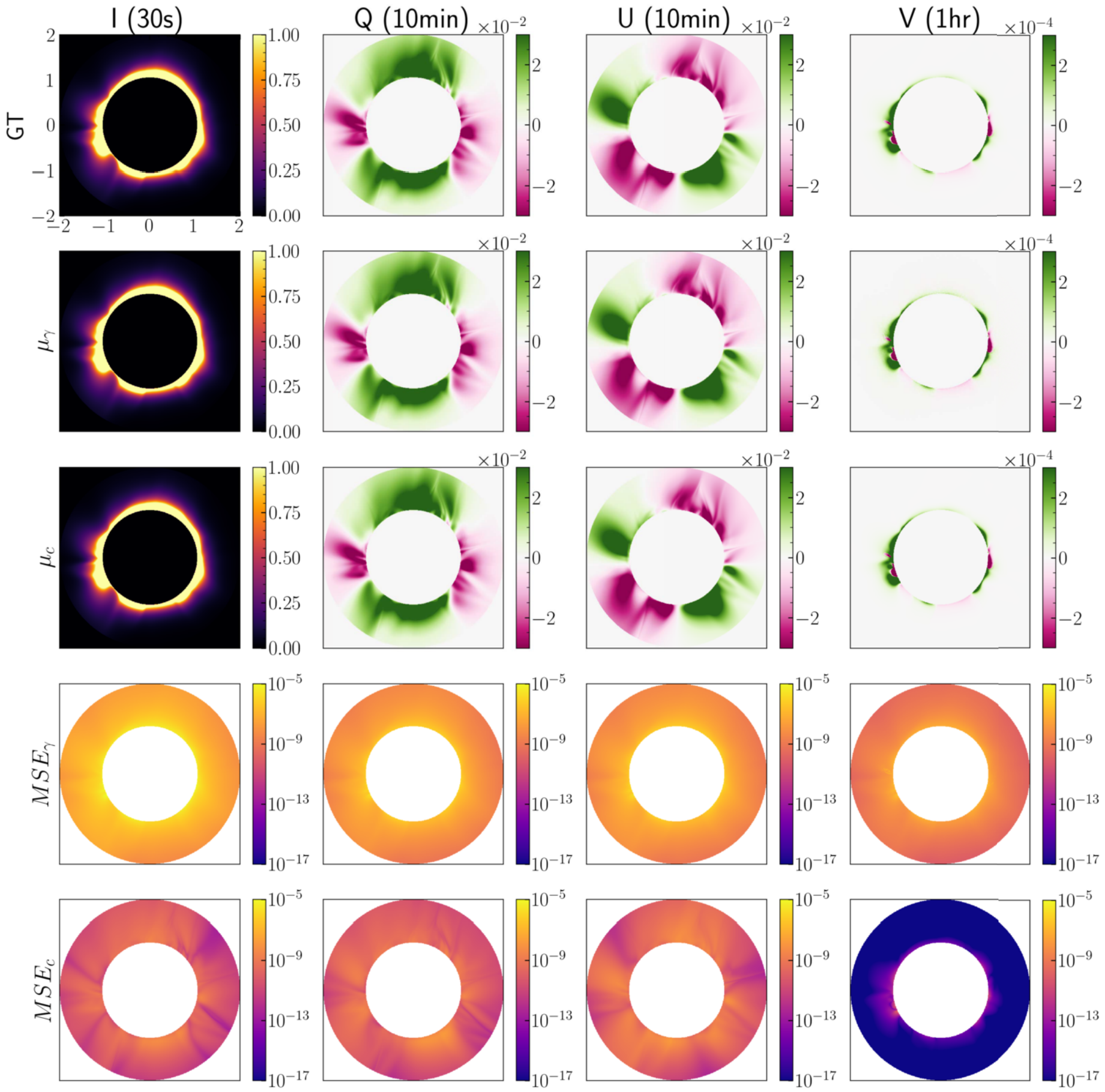

**Figure 7** Reconstruction of the PSI dataset of the solar corona, with Stokes vectors shown in each column. All pixels are normalized such that the maximum value of $I_{GT}$ is unity intensity so that relative comparison between panels can be easily made. The exposure times for mock observations are 30 seconds for $I$, 10 minutes for $Q$ and $U$, and 1 hour for $V$, which are approximately the cadences required for a reasonable signal. The ground truth $GT$ (first row) shows the raw data from the dataset. The second and third rows show the estimated mean of the reconstructed stokes vectors with only photon noise ($\mu_\gamma$) and calibration error ($\mu_c$), respectively. The mean squared error due to photon noise ($MSE_\gamma$) and that due to calibration error ($MSE_c$) are shown on the fourth and fifth rows with the same color scale, with yellow being the largest noise scale and violet being the smallest noise scale. We see that across all four Stokes elements, the photon noise MSE is on average larger than the calibration error MSE, indicating our calibration noise will not dominate the photon measurements.

Table 4 contains a summary of key statistical metrics used to evaluate the performance of our calibration reconstruction. We are ultimately interested in the contribution of calibra-tion uncertainty to the total uncertainty of calibration and photon noise for Stokes vector and b-field estimates. The first row of the table shows the pixel-wise average $\langle\rangle$ of $MSE_c$, and the second row shows the pixel-wise average due to photon noise $MSE_\gamma$. Noting that these two noise sources are independent, we compute the aggregate uncertainty as the sum $MSE_{c+\gamma} = MSE_c + MSE_\gamma$. The last row shows the contribution of $MSE_c$ to the total aggregate $MSE_{c+\gamma}$, respectively. We see that the contribution from calibration uncertainties is minimal in our expected photon noise level, which quantitatively indicates that predictions of the Stokes and b-field using posteriors calibrated by our Bayesian algorithm reach the photon-noise limited regime.

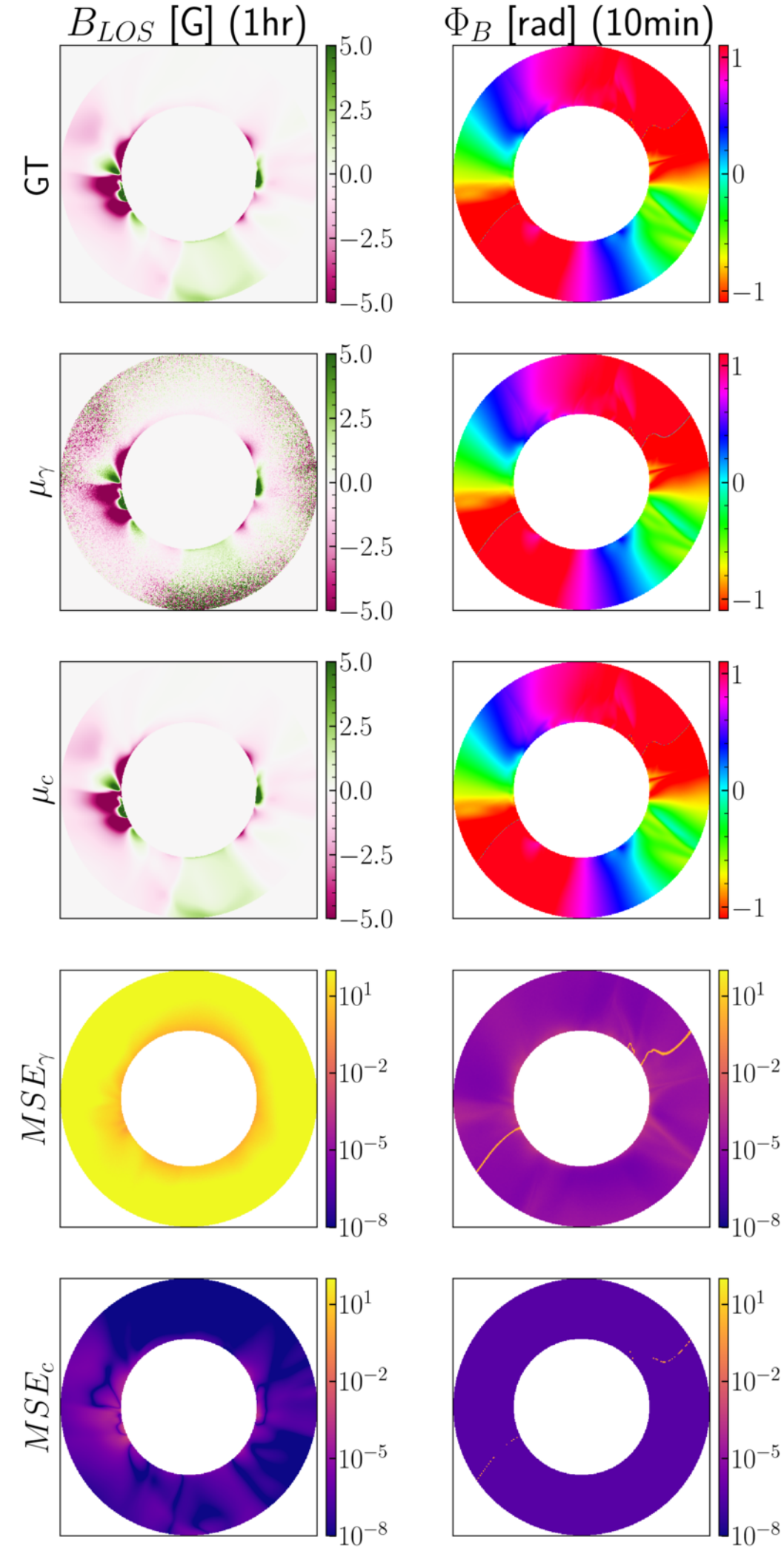

**Figure 8** Reconstruction of the coronal magnetic LOS intensity (left column) and the magnetic azimuth (right column). The ground truth (first row) shows the raw data computed from the PSI dataset using equations 5 and 6. The second and third rows show the mean of the reconstructed field with only photon noise ($\mu_\gamma$) and calibration error ($\mu_c$), respectively. The photon noise and calibration error MSE are shown on the fourth and fifth rows, respectively. We again observe that the photon noise dominates our calibration error MSE, which indicates that the probabilistic calibration produces a well-enough calibrated polarimeter.

**Table 4** Summary statistics for the performance of our calibration model on downstream Stokes elements (first 3 rows) and magnetic field estimates (last 3 rows), where ⟨⟩ denotes a 97% percentile pixel-wise average across the coronal pixels. The first row is the mean-squared error for the calibration noise estimate $MSE_c$, the second row shows the mean-squared error for the photon noise estimate $MSE_\gamma$, and the last row shows the fractional contribution of $MSE_c$ to the combined $MSE_{c+\gamma}$. We see that the overall contribution of the calibration uncertainty to the total uncertainty is very small. This shows that predictions of the Stokes vectors and b-field using posteriors calibrated by our Bayesian algorithm reach the photon-noise limited regime.

| | $I$ (30 s) | $Q$ (10 min) | $U$ (10 min) | $V$ (1 hr) |
|---|---|---|---|---|
| $\langle MSE_c \rangle$ | $1.54 \times 10^{-10}$ | $2.39 \times 10^{-10}$ | $5.44 \times 10^{-10}$ | $2.79 \times 10^{-16}$ |
| $\langle MSE_\gamma \rangle$ | $2.10 \times 10^{-7}$ | $3.13 \times 10^{-8}$ | $3.13 \times 10^{-8}$ | $3.56 \times 10^{-9}$ |
| $\langle MSE_c \rangle / \langle MSE_{c+\gamma} \rangle$ | $7.35 \times 10^{-4}$ | $7.59 \times 10^{-3}$ | $1.71 \times 10^{-2}$ | $7.84 \times 10^{-8}$ |
| | $B_{\text{LOS}}$ (1 hr) | $\Phi_B$ (10 min) | | |
| $\langle MSE_c \rangle$ | $4.67 \times 10^{-7}$ | $3.47 \times 10^{-7}$ | | |
| $\langle MSE_\gamma \rangle$ | $6.14 \times 10^{2}$ | $1.48 \times 10^{-5}$ | | |
| $\langle MSE_c \rangle / \langle MSE_{c+\gamma} \rangle$ | $7.61 \times 10^{-10}$ | $2.29 \times 10^{-2}$ | | |

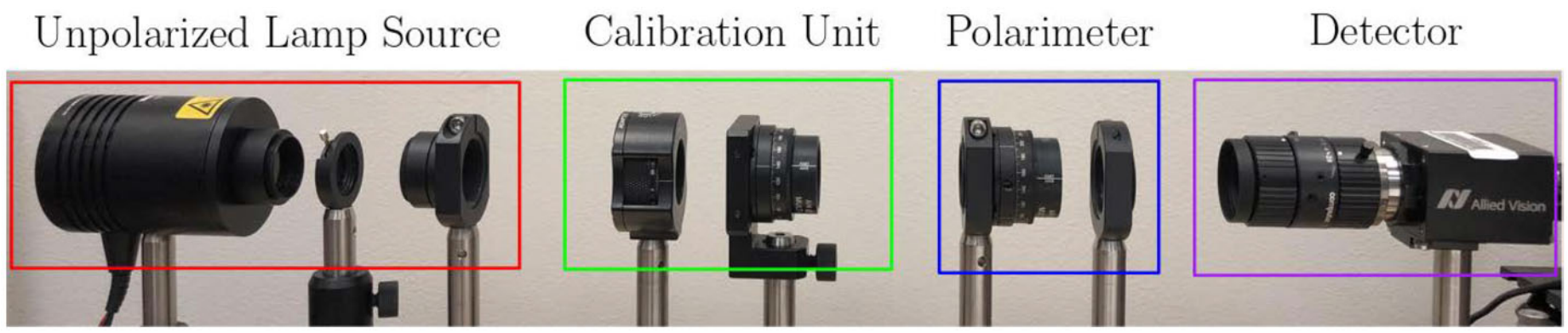

**Figure 9** Laboratory setup for the CORSAIR polarimeter calibration prototype. The unpolarized lamp source (red box) consists of a white light lamp, a 670 nm filter, an iris, and a 40 mm $CaF_2$ Plano-convex collimator. The Calibration Unit consists (green box) consists of a 400 - 700 nm linear polarizer, and a quarter waveplate at 670 nm. The Polarimeter (blue box) consists of a 630 nm quarter waveplate, and a 400 - 700 nm linear polarizer. Finally, the Detector (purple box) consists of an adjustable focus lens and the detector itself. Both waveplates and the Calibration Unit linear polarizer are placed on rotation stages to allow for any angular configuration in the system.

## 5. Lab Demonstration

### 5.1. Prototype Setup

While the previous section demonstrates that our calibration method is able to recover the parameters from a simulated polarimeter, it is equally important to demonstrate that the method is robust enough to recover parameters from actual measurements from a lab prototype with real-world uncertainties. We build a lab prototype to calibrate the CORSAIR polarimeter, and demonstrate that our probabilistic calibration method is able to recover both the parameters and the calibration uncertainty.

In Figure 9, we show a labeled image of our laboratory setup. We generate 670 nm collimated light from the lamp source (red box) using a broadband lamp, a filter, and a collimator. The light is then passed through a calibration unit (green box) consisting of a linear polarizer and a 670 nm quarter waveplate. Light from the calibration unit is then fed into the polarimeter (blue box) consisting of a 633 nm quarter waveplate and a linear polarizer (the analyzer). Because the actual CORSAIR waveplate retardance changes for the wavelengths planned to be measured, we choose to replicate this by using a 633 nm quartz multi-order quarter-waveplate: the retardance of the waveplate at 670 nm is approximately 2.434 radians predicted by the retardance equation $\phi(\lambda)$:

$$\phi(\lambda) = \frac{2\pi}{\lambda} \Delta n(\lambda) t, \tag{11}$$

where $\Delta n(\lambda)$ is the material-dependent birefringence computed using the Sellmeier equation, and $t$ is the optic thickness. To evaluate $\Delta n$, we used Sellmeier coefficients provided by Ghosh (1999).

Finally, we record the modulated intensity using a detector (purple box) consisting of a focus lens and a camera. Since we need access to all possible angle configurations of the calibration unit and polarimeter during data collection for calibration, we place both waveplates and the calibration unit linear polarizer in rotation mounts. As all polarization measurements are made with respect to the polarization axis of the analyzer, we align the other elements to the analyzer before proceeding with the calibration.

### 5.2. Experimental Results

Unlike calibration experiments using simulated data, laboratory experiments do not have access to the ground truth parameter values. Thus, to validate the performance of the calibration routine from laboratory measurements, we compare the recovered distribution with best theoretical estimates of the parameters. In particular, we make sure that the calibration is searching for offsets in the parameters that are larger than the various noise sources from lab measurements.

In our demonstration, we try to recover the waveplate retardances for both the calibration unit and polarimeter, and the intensity uncertainty from the measurements. Our best theoretical estimates for the retardances are $\phi^*_{\mathrm{CU}} = \pi/2$ for the calibration unit and $\phi^*_{\mathrm{Pol}} = 2.434$ for the polarimeter, and detector calibrations using 8-bit pixels indicate a noise level of around $\sigma^*_I = 0.02$ normalized to a reference polarimeter and calibration unit state. Our dataset consists of 100 samples of random configurations of the calibration unit and polarimeter and the corresponding output intensity, which we feed through our probabilistic calibration pipeline. We set our pipeline to have two unknown optical parameters, the waveplate retardance offset of the calibration unit and that of the polarimeter, and the $\sigma_I$ intensity uncertainty used in Equation 8. To show that our algorithm converges, we initialize the prior mean of these unknown parameters to values greater than the angular measurement noise of the prototype system.

In Figure 10, we show the resulting corner plot of the estimated joint posterior, with the kde estimates and 94% HDI intervals in blue on the diagonal. Table 5 compares the calibrated predictions with our theoretical guesses of the values. We see that the $\phi_{\mathrm{CU}}$ distribution agrees with the theoretical estimate, but our predicted $\phi_{\mathrm{Pol}}$ and $\sigma_I$ are different. We can expect our theoretical guess for $\phi_{\mathrm{CU}}$ to be fairly accurate, as it is a quarter waveplate made specifically for 670 nm. For $\phi_{\mathrm{Pol}}$, the systematic shift from the theoretical $\phi_S = 2.43$ radians to the inferred $\phi_{\mathrm{lab}} = 2.59$ radians is well-explained by the uncertainty in the theoretical retardance prediction, as shown in the error budget in the next section. Thus, we believe our Bayesian calibrated posterior is computing a more accurate estimate of the polarimeter waveplate retardance at 670 nm. Finally, while our predicted $\sigma_I$ was around 0.02 but the recovered one was 0.03, this makes sense, as our theoretical estimate does not account for angle measurement noise that can be amplified after passing it through the Muller matrix forward model. Thus, we show that our Bayesian calibration algorithm converges on real, noisy data collected from a lab prototype of the CORSAIR polarimeter.

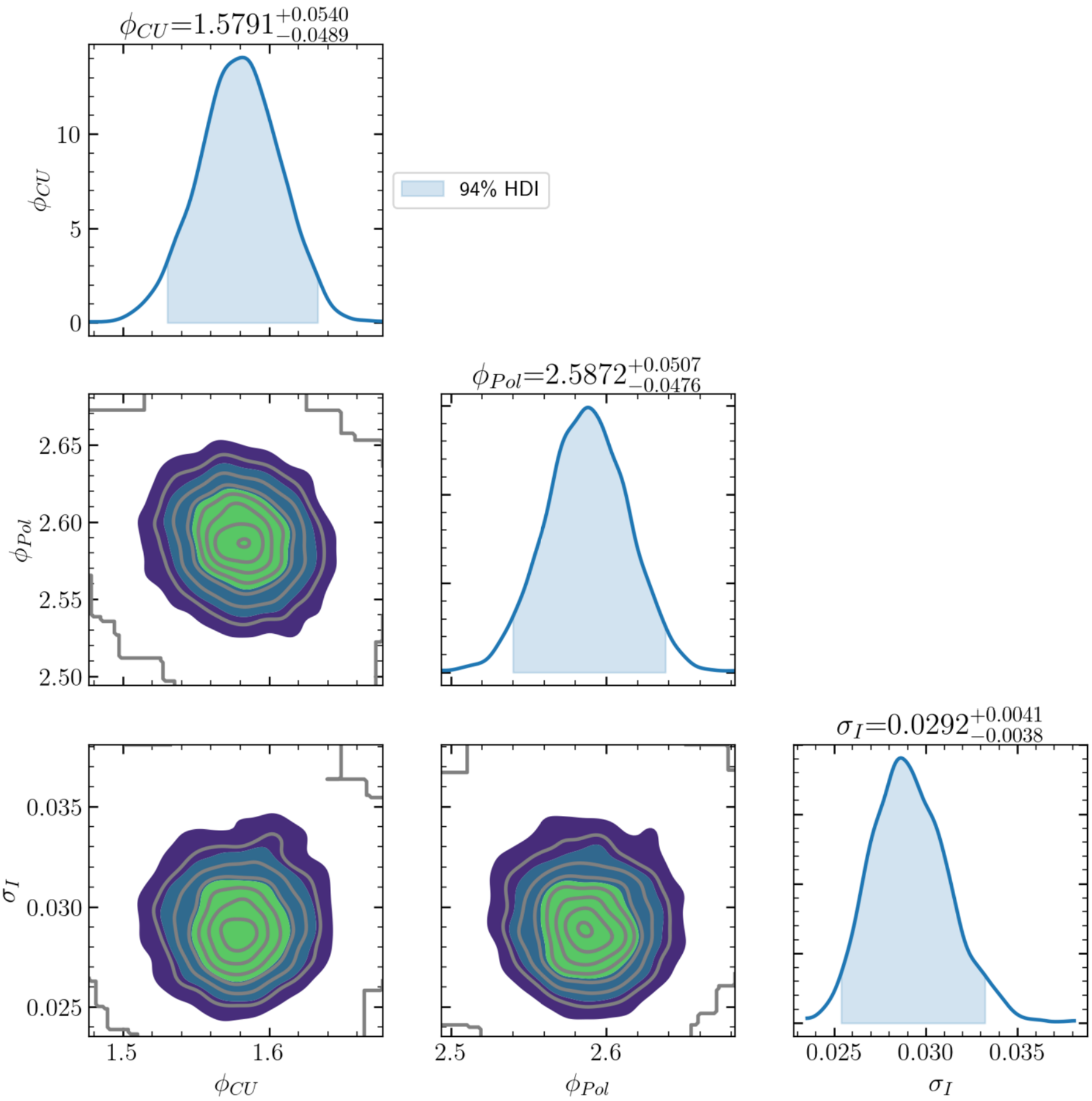

**Figure 10** Corner plot for a lab calibration experiment on the prototype model. The diagonal plots show the marginalized posterior density of the calibration unit and polarimeter unit waveplates $\phi_{\mathrm{CU}}$ and $\phi_{\mathrm{Pol}}$, and the estimated intensity uncertainty $\sigma_I$, with the 94% Highest Density Interval (shaded blue). The off-diagonal plots show the 2D projected posterior distributions, using Hamiltonian Monte Carlo NUTS samples to estimate the posterior density. There is close agreement between the calibrated estimates and our theoretical guesses for these parameters.

**Table 5** Summary table of estimates and theoretical parameter values for the lab demonstration. The Lab column indicates the estimated mean and 94% HDI values. The Theory column indicates the theoretical estimates for each parameter value. We see that the $\phi_{\mathrm{CU}}$ estimate matches our guess, but the other two do not. We believe that the theoretical estimate for $\phi_{\mathrm{Pol}}$ using the retardance equation may not appropriately capture the true underlying birefringence of the waveplate, and that our estimate of the intensity uncertainty does not include angle measurement noise that could be amplified through the forward Mueller matrix model.

| Param | Lab | Theory |
| --- | --- | --- |
| $\phi_{\mathrm{CU}}$ | $1.5791^{+0.0540}_{-0.0489}$ | 1.5708 |
| $\phi_{\mathrm{Pol}}$ | $2.5872^{+0.0507}_{-0.0476}$ | 2.4338 |
| $\sigma_I$ | $0.0292^{+0.0041}_{-0.0038}$ | 0.02 |

**Table 6** Error budget summary for the waveplate retardances. Four major contributions and the RSS total are provided by the rows for both $\phi_{\rm CU}$ and $\phi_{\rm Pol}$. Columns show both the native uncertainty in units of the noise sources, and the corresponding propagated uncertainty in the retardance space. For $\phi_{\rm CU}$, the dominant source is the vendor retardance tolerance at 0.06 radians, while for $\phi_{\rm Pol}$ the dominant source is the filter bandwidth uncertainty at 0.390 radians. We find that the RSS error for $\phi_{\rm CU}$ is consistent with the posterior predicted uncertainty, and that the RSS error for $\phi_{\rm Pol}$ explains the systematic offset in the posterior mean relative to the theoretical estimate.

| | $\phi_{\rm CU}$ | | $\phi_{\rm Pol}$ | |
|---|---|---|---|---|
| | Native uncertainty | $\sigma_\phi$ | Native uncertainty | $\sigma_\phi$ |
| Angle Setting | 0.5° | 0.006 rad | 0.5° | 0.006 rad |
| Vendor Tolerance | – | 0.063 rad | – | – |
| Sellmeier Model | – | – | $10^{-6}$ | 0.009 rad |
| Filter Bandwidth | 10 nm | 0.007 rad | 10 nm | 0.390 rad |
| Total (RSS) | – | 0.064 rad | – | 0.390 rad |

## 5.3. Error Budget

To further interpret the recovered waveplate retardances from the lab calibration, we construct an error budget summarizing the dominant sources of uncertainty in the prototype setup. Table 6 shows the error budget for four main sources of noise in the rows: the angle setting, vendor tolerance, theoretical Sellmeier model, and filter bandwidth. For each of the two parameters, we show both the native uncertainty in the units of the noise source, and uncertainty $\sigma_\phi$ propagated to retardance space.

The angle setting was done by hand using rotation stages with 1° tick marks. Therefore, we take the angle uncertainty to be 0.5°. Propagating this through using MCMC samples results in a retardance uncertainty of about 0.006 radians for both $\phi_{\rm CU}$ and $\phi_{\rm Pol}$. Likewise, the vendor tolerance of the CU waveplate is $\lambda/100$: converting this to retardance space by multiplying by $2\pi/\lambda$ gives $2\pi/100$, or around 0.06 radians, which is an order of magnitude larger than the angular setting contribution. For the Pol waveplate, since we are not operating at the manufacturer-specified wavelength, we compute the error on the birefringence using published comparisons between experimental and calculated values. For quartz, values indicated by Table 8 of (Ghosh 1999) give an uncertainty of $10^{-6}$ at 670 nm, and propagating this through the retardance equation gives $\sigma_\phi \approx 0.009$ radians.

The filter bandwidth is reported to be 10 nm by the manufacturer, or ±5 nm. We compute $\sigma_\phi$ by using a theoretical error propagation through the retardance equation (see Appendix E for more details). For $\phi_{\rm CU}$, this amounts to 0.007 radians, which is of the same order as the angular setting contribution. However, for $\phi_{\rm Pol}$ the error gives 0.390 radians: the polarimeter waveplate is many-waves thick at 670 nm, and thus is sensitive to a small change in the wavelength in the range of the filter bandwidth.

Finally, we also considered the error from lamp jitter. A live-feed of the source intensity during calibration indicates no change in an 8-bit quantization of the lamp measurements, which translates to a maximum ADC step error of $\Delta = 1/255$ when pixels are normalized to $[0, 1]$. Assuming uniform distribution for the jitter over the range $[-\Delta/2, +\Delta/2]$, the uncertainty in intensity space is $\Delta/\sqrt{12}$, or in our case around 0.0011 in normalized intensity units. Since our posterior predicts a $\sigma_I \approx 0.03$ which is over an order of magnitude larger, we conclude that quantization is not a limiting noise source, and do not include it in the RSS error budget and let the posterior $\sigma_I$ be an effective intensity-noise term that includes this effect.

We combine the dominant four error contributions in the last row using a root sum of squares (RSS), and find that the dominant source of error is the vendor tolerance for $\phi_{\mathrm{CU}}$, and the RSS is 0.064 radians, which is consistent with the posterior predicted uncertainty around 0.05 radians. For $\phi_{\mathrm{Pol}}$, we compute an RSS of 0.390 radians, with the dominant contribution from the filter bandwidth uncertainty. To check if this theoretical prediction $\sigma_{\phi,\mathrm{RSS}} = 0.390$ radians is consistent with our experimental posterior uncertainty $\sigma_{\phi,\mathrm{post}} \approx 0.05$ radians, we assume independence in these effects and compute the combined uncertainty $\sigma_{\mathrm{comb}} = \sqrt{\sigma_{\phi,\mathrm{RSS}}^2 + \sigma_{\phi,\mathrm{post}}^2}$, and a corresponding z-score for the systematic shift we find from the lab calibration:

$$z = \frac{\Delta\phi_{\mathrm{Pol}}}{\sigma_{\mathrm{comb}}}. \tag{12}$$

Taking $\Delta\phi_{\mathrm{Pol}} \approx 0.1$ radians as the systematic shift computed from Table 5, and computing $\sigma_{\mathrm{comb}} = 0.393$ radians, we get a z-score of 0.42, indicating the measured systematic difference is well-within the combined uncertainty due the error budget RSS and the posterior. We therefore treat the 0.1 radians as a systematic offset that is explained by the RSS error budget.

## 6. Conclusion

We present a novel Bayesian model for spectropolarimeters and a corresponding probabilistic calibration algorithm for the CORSAIR polarimeter that is able to propagate calibration uncertainties through posterior distributions of calibration parameters, Stokes measurements, and magnetic field measurements. The probabilistic model combines existing Mueller matrix representations of polarimeters with probabilistic parameters that represent physical imperfections we believe to be present in the actual system.

To calibrate the system, we place a calibration unit in front of the polarimeter to generate known polarized light, and collect data consisting of output intensities from the calibration unit and polarimeter rotated to different states. The calibration process modifies the initial probability distribution of the parameters (prior) into a more likely distribution (posterior) based on the collected data. The posterior distribution captures the complex uncertainties and covariances between the different parameters that would otherwise be very hard to explicitly model.

We test the accuracy and convergence of our calibration algorithm by using a simulated polarimeter and calibration unit to collect data. We find that our Bayesian calibration algorithm is able to converge, and produces a well-constrained posterior over the parameters, with uncertainties consistent with ground truth. Posterior predictive checks are done to further validate that the generative model can reproduce the calibration data within the expected predictive uncertainty, and learn the dominant sources of systematic effects.

We further show that the calibration algorithm enables us to propagate the calibration error and uncertainties to downstream predictions. Using the posterior distribution of the parameters, we are able to compute the distribution of demodulation matrix entries, which allows us to further reconstruct the Stokes and the magnetic field LOS strength and POS azimuth. We test the uncertainty propagation by simulating spectropolarimetric observations from a magnetohydrodynamic model of the solar corona. We find that the contribution of calibration uncertainty towards the reconstructed results is minimal relative to that of the photon noise uncertainty, indicating that estimates using our Bayesian calibration are within the photon-noise-limited regime.

Finally, we test the convergence of our calibration algorithm on lab-collected data from a CORSAIR prototype. Our algorithm is able to converge to a posterior distribution that has peaks close to theoretical estimates of the parameters, demonstrating that this algorithm works on real data. We further validate the statistical consistency of our posterior distribution by comparing it to a concise error budget that theoretically predicts the dominant sources of noise.

Future work will be done to refine both the probabilistic model and the calibration process. We plan to add additional parameters, such as diattenuation and birefringence terms directly into the Mueller matrix model that more accurately reflect optical imperfections within the system. Likewise, we plan to use more sophisticated and efficient sampling processes to estimate the parameter posterior. We also plan to further study the combination of both photon noise uncertainty and calibration uncertainty in downstream magnetic field measurements, as opposed to comparing them separately, and refining our calibration algorithm so that the final reconstructed magnetic field measurements are photon-noise-limited. Additionally, these magnetic field reconstructions rely on the POS approximation and do not marginalize over line-of-sight integration effects that smooth the complex 3D magnetic field geometries. Future work can be done to address the systematic biases for LOS-generated coronal emission by incorporating a forward model of coronal polarization into the Bayesian inference pipeline. Finally, we will apply and refine this calibration procedure towards the actual CORSAIR polarimeter once its construction is completed.

## Appendix A: UCoMP Calibration Algorithm

The UCoMP calibration algorithm was designed by Steven Tomczyk, and focuses on recovering the demodulation matrix of the polarimeter by performing a boot-strapped optimization process to solve for the calibration inputs. First, data is collected and stored. For each calibration input generated from the calibration optics, we modulate the polarimeter to through $n_{\text{mod}}$ states to produce varying output intensities. With $n_{\text{cal}}$ different inputs, our data can be arranged in a $I_{\text{meas}}$ matrix that is $n_{\text{mod}} \times n_{\text{cal}}$, where each column are the modulated intensities for each calibration input.

In Figure 11, we show a schematic diagram of the iterative calibration process. The input unpolarized light is modeled as a Stokes vector $S_c$, which in theory should be $[1, 0, 0, 0]^T$ but in reality contains imperfections from the diffuse source. With assumptions about the calibration optics, we can then compute the $4 \times n_{\text{cal}}$ calibration matrix $C$, where column $j$ is the calibration input into the polarimeter that produces the modulated intensities in column $j$ of $I_{\text{meas}}$. From $C$ we can compute the demoulation matrix of the polarimeter $D$, which we then use to back-compute the $4 \times n_{\text{cal}}$ calibration matrix labeled as $S_{cal}$.

The calibration process minimizes the root-mean-square error between forward model calibration matrix $C$, and the inverse model calibration matrix $S_{\text{cal}}$. The forward model is effectively the theoretical model of the calibration optics that produces $C$ from an input light source. The inverse model utilizes both $C$ from the forward pass and the calibration data $I_{\text{meas}}$ to back-compute $C$ which we will call $S_{\text{cal}}$. The demodulation matrix $D$, which characterizes the modulation states of the polarimeter, is computed in an intermediate step during the inverse model.

The state space is a 6-dimensional vector $\theta = [Q, U, V, \phi, T, \delta\phi]$, where $[Q, U, V]$ are the Stokes values of $S_c$, $\phi$ is the retardance angle of the calibration retarder, $T$ is the transmission of the calibration optics, and $\delta\phi$ is the retarder offset. In other words, the calibration

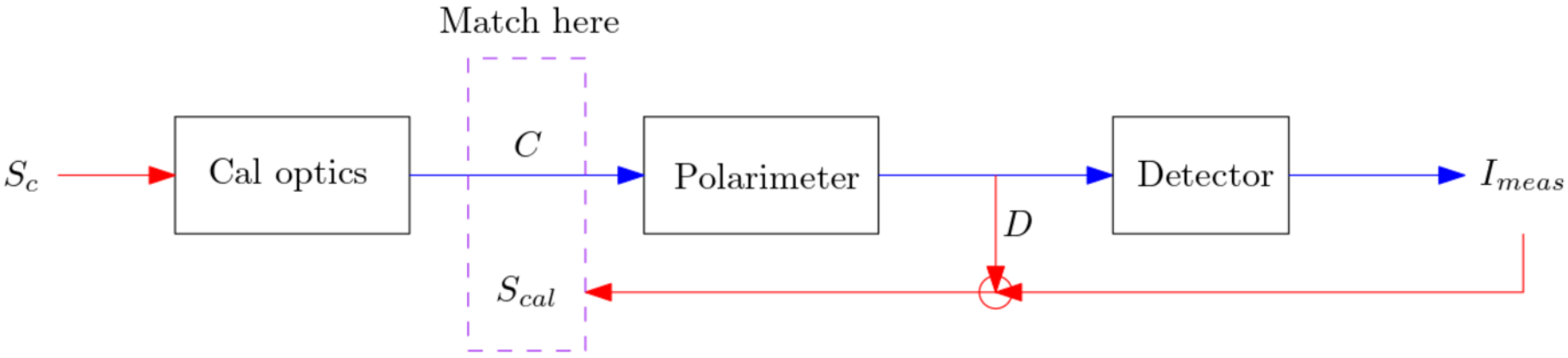

**Figure 11** Schematic diagram of the UCoMP calibration process.

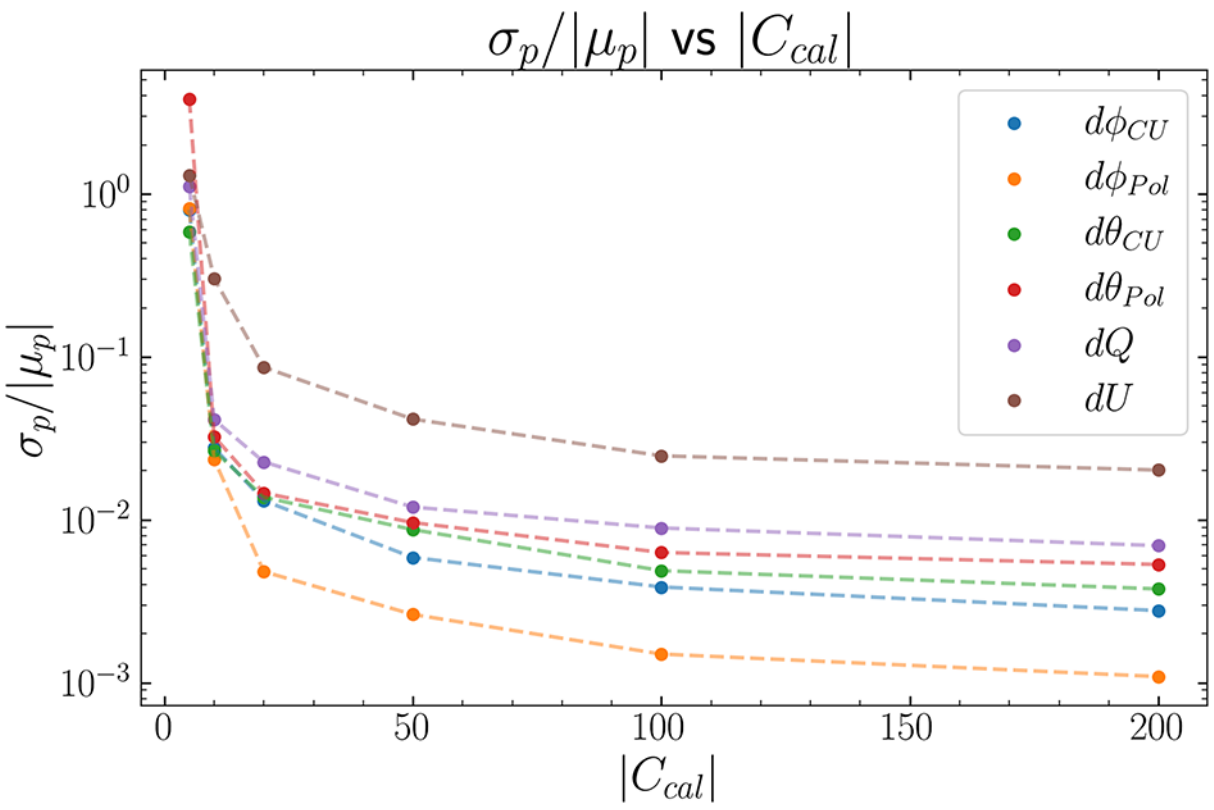

**Figure 12** Plot of $\sigma_p/|\mu_p|$, the uncertainty-to-mean ratio, for each parameter as a function of $|C_{\text{cal}}|$, the number of calibration data samples. We see that the calibration precision converges at around 100 samples, but the behavior is well-captured even at 50 samples.

process can be formulated as an optimization problem given by

$$\theta^* = \min_{\theta} \|C - S_{\text{cal}}\|^2,$$

and specifically for UCoMP the algorithm uses a Powell Minimization scheme to solve for the optimal $\theta^*$. In our results in Figure 3, we use a derivative-free optimization algorithm called the Covariance Matrix Adaptation Evolution Strategy (e.g. Hansen, Müller, and Koumoutsakos 2003; Igel, Suttorp, and Hansen 2006; Hansen and Ros 2010), but otherwise utilize the same optimization strategy as described in this section as the baseline point estimates of the parameters.

## Appendix B: Fractional Uncertainty of Parameters

To determine the number of calibration samples needed for a reasonable probabilistic fit, we test the Bayesian routine with our simulation pipeline with different amounts of calibration data $|C_{\text{cal}}|$, and plot the fractional uncertainty for each parameter. The fractional uncertainty is the posterior standard deviation divided by the posterior mean with magnitude $\sigma_p/|\mu_p|$, and it is a dimensionless measure of the calibration precision for each parameter.

Figure 12 shows the results for all 6 parameters. We see that most of the convergence behavior is captured within 50 samples, and after 100 samples the change is almost negligible. Note that the lamp polarization offsets $dQ$ and $dU$ have a high error: this is expected, as they have the weakest identifiability among the 6 parameters, but at 50 samples the largest fractional change ($dU$) is already with 5%. In Section 4, we will choose to use 50 calibration data samples for our analysis: this captures most of the convergence while also being computationally efficient for downstream tasks.

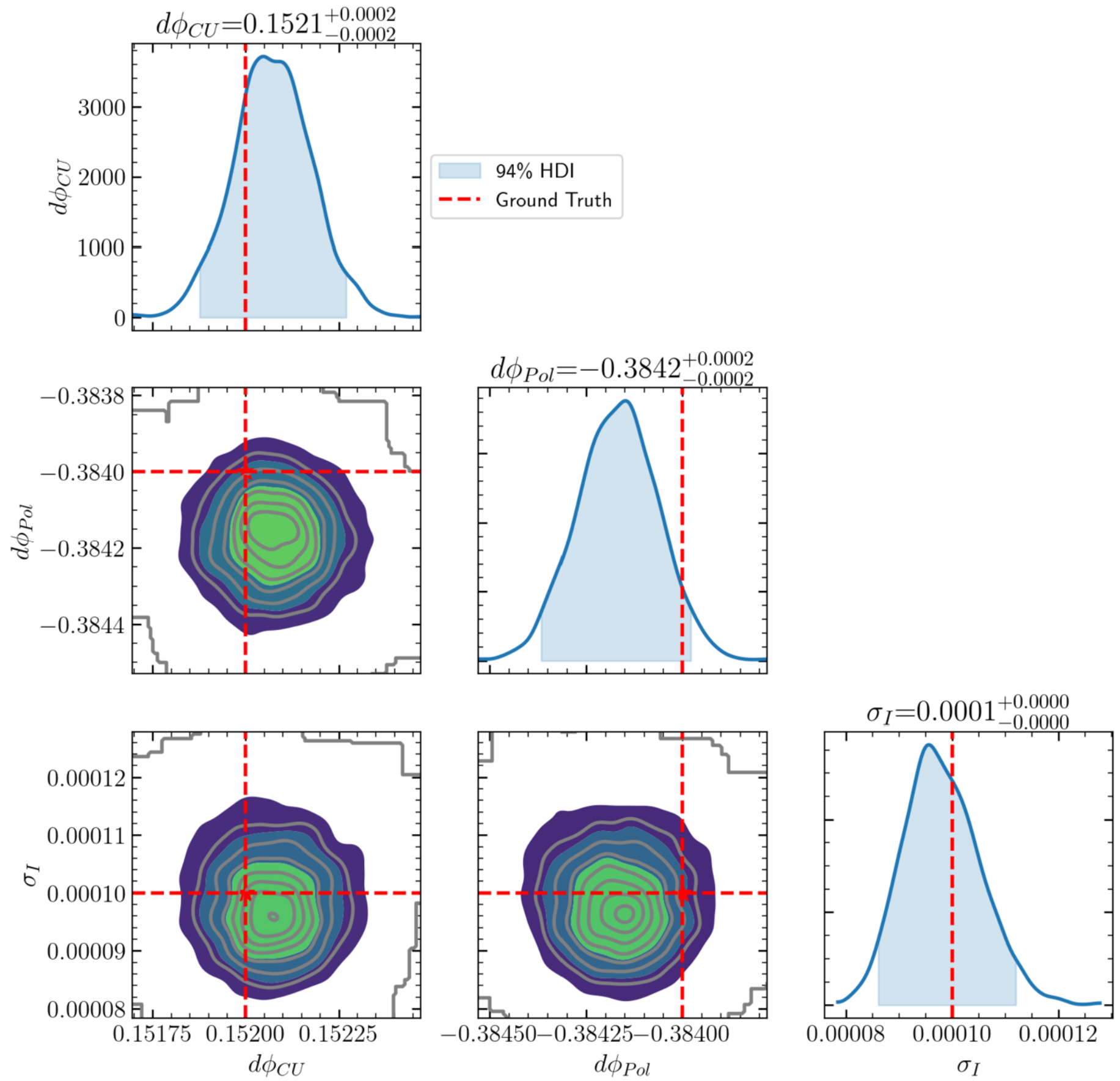

**Figure 13** Corner plot for a calibration experiment on the simulated CORSAIR model with only photon noise. The diagonal plots show the marginalized posterior density (blue) of the 2 optical parameters $d\phi_{\mathrm{Pol}}$ and $d\phi_{\mathrm{CU}}$ and the estimated intensity uncertainty $\sigma_I$, with the ground truth value (dashed red) and the 94% Highest Density Interval (shaded blue). The off-diagonal plots show the 2D projected posterior distributions, using Hamiltonian Monte Carlo NUTS samples to estimate the posterior density. We see that that the calibration model accurately captures the ground truth values of all 3 parameters to the shaded credible interval. This experiment demonstrates that the Bayesian calibration algorithm can correctly recover the intensity uncertainty from simulated data.

## Appendix C: Photon Noise Experiments

To test whether or not our Bayesian model can learn the intensity uncertainty $\sigma_I$ (Equation 8), we set all other sources of measurement noise added throughout the simulator to zero and only add photon noise to the observed modulated intensities. The $\sigma_I$ posterior recovered by the Bayesian calibration algorithm is thus an estimate of the photon noise we add. In our experiment, we calibrate the system with three free parameters: $d\phi_{\mathrm{Pol}}$ and $d\phi_{\mathrm{CU}}$ retardances (see Table 1 for details), and $\sigma_I$. We set the Gaussian noise to be 0.0001, or a SNR of about 10000, which we can expect to reach with a reasonable exposure time in the lab. In Figure 13, we show the resulting corner plot for the recovered posterior with 3 free parameters: we see that the HDI interval (shaded blue) of the posterior on the diagonal plots contains the ground truth (dashed red) values. More importantly, this experiment shows the ability for the Bayesian calibration algorithm to correctly recover $\sigma_I$, which is helpful in scenarios where we cannot fully forward model the noise.

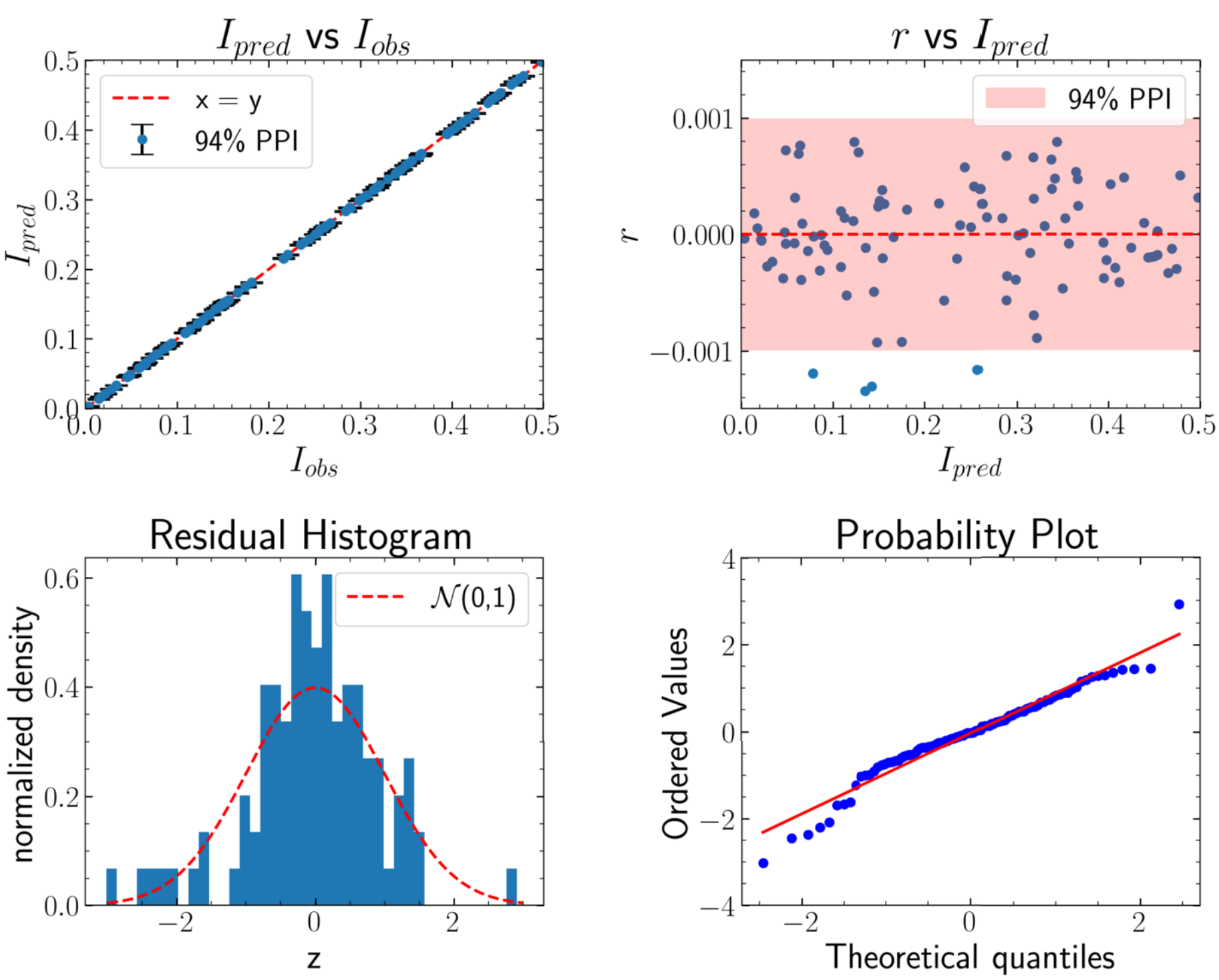

**Figure 14** Posterior predictive checks for the calibrated posterior. The predicted-vs-observed intensity plot (top left) shows that the generative model and inferred parameters can reproduce the calibration data within the expected predictive uncertainty. The residuals plot (top right) shows that there is no structure or intensity dependence of the residuals, indicating that the calibration model learns the dominant systematic effects. The residual histogram (bottom left) shows that the normalized residuals are approximately centered at zero and consistent with a standard normal distribution. The QQ plot (bottom right) shows that the normalized residuals are consistent with the Gaussian noise model assumption, and the slight deviation at the tails are consistent with the finite-sample variability of 100 calibration measurements.

## Appendix D: Additional Posterior Predictive Checks

In this section, we provide two additional posterior predictive checks along with the $I_{\mathrm{pred}}$ vs $I_{\mathrm{obs}}$ and residual plots shown in Figure 4. Together, these diagnostics comprehensively validate the predictive data reproduction, residual structure, and noise distribution of the posterior predictive abilities of the calibrated Bayesian model.

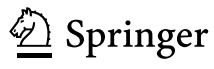

In Figure 14, we show 4 plots that validate the correctness of the calibrated posterior, using a larger 100 sample calibration dataset rather than the 50 used in the analysis. For this larger dataset, we observe a similar behavior in the top row plots: the $I_{\text{pred}}$ vs $I_{\text{obs}}$ close agreement indicates that generative model is able to reproduce the high SNR calibration measurements within the expected predictive uncertainty, and the structureless residuals indicate that the generative model is able to capture the dominant systematic effects in the calibration procedure. The bottom left panel overlays a standard normal distribution on the histogram of the residuals, which provides a complementary view of the residuals, showing that the normalized residuals are approximately centered at zero and consistent with a standard normal distribution. Finally, we show a quantile-quantile (QQ) plot in the bottom right panel. For most of the range, the normalized residuals are consistent with the Gaussian noise assumption, with slight deviations in the tails that are consistent with finite-sample variability of 100 calibration measurements.

Finally, as an additional quantitative check of the likelihood model, we compute the RMS of the calibration residuals $r_i = I_{\text{obs},i} - I_{\text{pred},i}$ normalized by the inferred noise parameter $\sigma_I$:

$$\frac{RMS}{\sigma_I} = \frac{\sqrt{\frac{1}{N}\sum r_i^2}}{\sigma_I} \tag{13}$$

We expect the $RMS/\sigma_I$ to be close to unity if the assumed Gaussian noise model adequately describes the data. Indeed, we find that $RMS/\sigma_I \approx 0.96$: this means that the observed residual scatter is consistent with the noise level implied by the likelihood model.

## Appendix E: Filter Bandwidth Error Propagation

To determine the error $\sigma_\phi$ due to filter bandwidth $\sigma_\lambda$, we will assume small uncertainties and propagate error through a linearization of $\phi(\lambda)$:

$$\sigma_\phi \approx \left|\frac{\partial \phi}{\partial \lambda}\right| \sigma_\lambda. \tag{14}$$

Differentiating Equation 11 and substituting into Equation 14 gives

$$\sigma_\phi \approx \sigma_\lambda \left| -\frac{\phi(\lambda)}{\lambda} + \frac{2\pi t}{\lambda}\frac{d(\Delta n(\lambda))}{d\lambda} \right|. \tag{15}$$

In our prototype setup, the filter bandwidth is 10 nm, and assuming a uniform distribution over this range, the uncertainty in $\lambda$ is given by $\sigma_\lambda = 10/\sqrt{12}$ nm.

For $\phi_{\text{CU}}$, we assume that since the optic is a quarter waveplate operating at the designed wavelength 670 nm, the deviations due to the birefringence is negligible compared to the leading order term $\phi/\lambda$. Thus, with $\sigma_\lambda = 10/\sqrt{12}$ nm and $\phi(670\text{ nm}) = \pi/2$ radians, we get an uncertainty of $\sigma_\phi = 0.007$ radians.

For $\phi_{\text{Pol}}$, we evaluate both terms in Equation 15 since the birefringence may change significantly at 670 nm since the quarter waveplate was originally designed to operate at 633 nm. The birefringence term $\Delta n = |n_e - n_o|$ is computed using empirically tabulated ordinary $n_o$ and extraordinary $n_e$ refractive indices of Crystalline Quartz at 670 nm found on the Refractive Index Database (Ghosh 1999). We evaluate Equation 15 with $\sigma_\lambda = 10/\sqrt{12}$, $\phi(670\text{ nm}) \approx 2.434$, and $t = 996.11$ microns, and numerically evaluate the derivative of $\Delta n$. We get that the derivative terms is negligible compared to the leading order term $\phi/\lambda$, and get an uncertainty of $\sigma_{\phi,\text{Pol}} = 0.390$ radians.

**Acknowledgments** Support for Alan Hsu is provided by a grant from the Brinson Foundation. CORSAIR is supported by NASA awards 80NSSC21K0809 and 80NSSC26M0007. This paper is a direct extension on the previous work done in Hsu et al. (2025).

**Author contributions** A.H. wrote the full paper, including the text, figures, references, and appendices, with all other authors contributing by reviewing and providing suggestions to various parts of the manuscript. J.S. provided the main guidance for Alan Hsu throughout the research, and largely contributed to the structure of the paper, the scientific context on which the research is conducted, and the implications of having computed the calibration uncertainty. S.T. provided the background to existing solar polarimeter calibration methods, and particularly helped with the technical aspects of Appendix A (UCoMP calibration). M.K. generated the LOS-integrated polarization maps plotted in Figure 5. This data was also used in Section 4 to validate the reconstruction of the magnetic field values.

**Data Availability** No datasets were generated or analysed during the current study.

## Declarations

**Competing interests** The authors declare no competing interests.